\documentclass[11pt, a4paper]{article}
\pdfoutput=1

\usepackage{jcappub}

\usepackage{comment}
\usepackage{amsmath}
\usepackage{amsfonts}
\usepackage{amssymb}
\usepackage{graphicx}
\usepackage{mathrsfs}
\usepackage{latexsym}
\usepackage{color}
\usepackage{slashed, cancel}
\usepackage{hyperref}
\usepackage{tikz}
\usepackage{bbold}
\usepackage{soul} 
\usepackage{mathtools}
\newcommand{\fsl}[1]{\ensuremath{\mathrlap{\!\not{\phantom{#1}}}#1}}
\usepackage[font=footnotesize,labelfont=bf]{caption} 
\usepackage{subcaption}
\usepackage{enumitem}
\usepackage{journals}
\allowdisplaybreaks

\hypersetup{urlcolor=blue, colorlinks=true} 

\usepackage{fancyhdr}
\numberwithin{equation}{section}

\definecolor{rossos}{rgb}{0.8,0.2,0.3}
\definecolor{bluscuro}{rgb}{0.15, 0.2, .85}
\definecolor{bluchiaro}{cmyk}{1,.3,0.,0.1}
\hypersetup{colorlinks, citecolor=bluscuro, linkcolor=bluscuro, urlcolor=bluscuro}

\makeatother   
\newcommand{\GeV}{{\rm \,GeV}}
\newcommand{\TeV}{{\rm TeV}}

\newcommand{\cm}{{\rm \,cm}}

\newcommand{\mathsc}[1]{\text{\textsc{#1}}}

\newcommand{\met}{\slashed{E}_T}

 \def\be   {\begin{equation}}   \def\ee   {\end{equation}}
 \def\ba   {\begin{array}}      \def\ea   {\end{array}}
 \def\bea  {\begin{eqnarray}}   \def\eea  {\end{eqnarray}}
 \def\bean {\begin{eqnarray*}}  \def\eean {\end{eqnarray*}}
 \def\nn{\nonumber}

\begin{document}

\today

\title{Self-consistent Dark Matter Simplified Models with an s-channel scalar mediator}

\author{Nicole F.\ Bell,}
\author{Giorgio Busoni and}
\author{Isaac W. Sanderson}
\affiliation{ARC Centre of Excellence for Particle Physics at the Terascale \\
School of Physics, The University of Melbourne, Victoria 3010, Australia}

\emailAdd{\tt n.bell@unimelb.edu.au}
\emailAdd{\tt giorgio.busoni@unimelb.edu.au}
\emailAdd{\tt isanderson@student.unimelb.edu.au}

\abstract{We examine Simplified Models in which fermionic DM interacts
  with Standard Model (SM) fermions via the exchange of an $s$-channel
  scalar mediator.  The single-mediator version of this model is not
  gauge invariant, and instead we must consider models with two scalar
  mediators which mix and interfere.  The minimal gauge invariant
  scenario involves the mixing of a new singlet scalar with the
  Standard Model Higgs boson, and is tightly constrained.  We
  construct two Higgs doublet model (2HDM) extensions of this
  scenario, where the singlet mixes with the 2nd Higgs doublet.
  Compared with the one doublet model, this provides greater freedom
  for the masses and mixing angle of the scalar mediators, and their
  coupling to SM fermions.  We outline constraints on these models,
  and discuss Yukawa structures that allow enhanced couplings, yet
  keep potentially dangerous flavour violating processes under control.
  We examine the direct detection phenomenology of these models,
  accounting for interference of the scalar mediators, and
  interference of different quarks in the nucleus.  Regions of
  parameter space consistent with direct detection measurements are
  determined.}

\maketitle


\section{Introduction}

The quest to uncover the identity of dark matter (DM) is one of the
key challenges facing fundamental physics.  Of the many types of dark
matter candidates proposed, Weakly Interacting Massive Particles
(WIMPs) are singled out as theoretically compelling candidates that
can be meaningfully probed in current and forthcoming experiments.
Indeed, the absence of a dark matter signal in the LHC run I data,
together with complementary constraints from direct and indirect
detection searches, has already excluded a non-trivial portion of the
WIMP parameter space.  The dark matter community is thus eagerly
awaiting new data pertaining to the higher energy collisions of the
LHC run II.

In order to interpret the results of these experiments, it is crucial
to have a suitable framework to describe DM interactions with Standard
Model (SM) particles.  To this end, Effective Field Theories (EFTs)
were used in many of the LHC 8 TeV analyses. However, while an EFT
approach is valid for low momentum transfer processes such in dark
matter direct detection, it will break down at energy scales
comparable to or larger than that of the new physics.  Their use for
LHC dark matter searches is thus not optimal, given the WIMP energy
scale is expected to be comparable to the EW scale and hence directly
accessible at the LHC. Indeed current LHC limits on the EFT energy
scale fall in the range $\Lambda \sim O(\textrm{GeV-TeV})$
\citep{Busoni:2013lha,Busoni:2014sya,Busoni:2014haa,Buchmueller:2013dya,Abdallah:2014hon,Shoemaker:2011vi}.

To remedy this issue, Simplified Models have been developed as a
superior alternative that retain some of the desirable features of the
EFTs (simple generic descriptions that approximate the relevant
phenomenology of a broad spectrum of UV complete models) while
remaining valid at higher energy scales.  They do this through the
explicit introduction of a particle that mediates the interactions of
DM with SM fermions.  The benchmark simplified models, outlined in the
report of the LHC DM Forum (DMF), involve fermionic dark matter
interacting via the exchange of a neutral spin-0 $s$-channel mediator,
a neutral spin-1 $s$-channel mediator, or a charged spin-0 $t$-channel
mediator
\citep{Shoemaker:2011vi,Abdallah:2014hon,Abdallah:2015ter,Abercrombie:2015wmb,Boveia:2016mrp,Jacques:2016dqz,Buckley:2014fba,Harris:2014hga}.

While these simplified models are a definite improvement over the EFT
approach, they are still not ideal.  Indeed, some of the benchmark
simplified models outlined by the DMF are clearly inadequate, as they
are not gauge invariant and hence are not renormalizable.  Issues
associated with the lack of gauge invariance (with respect to both SM
and dark-sector gauge groups) have been discussed in
\citep{Bell:2015sza,Bell:2015rdw,Haisch:2016usn,Englert:2016joy,Kahlhoefer:2015bea,Bell:2016fqf,Ko:2016zxg}.
For example, models that involve the exchange of a spin-1 mediator
($Z'$) with axial-vector couplings to fermions is not gauge invariant
unless a dark-Higgs is introduced to unitarize the longitudinal
component of the $Z'$
\citep{Kahlhoefer:2015bea,Bell:2016fqf,Duerr:2016tmh,Bell:2016uhg}. The
minimal self-consistent scenario involving an axial vector mediator
must then be expanded to include {\it two} mediators -- the $Z'$ and
scalar.  The single mediator benchmark simplified models do not
accurately capture the phenomenology of such two mediator models.

In this paper we consider the case of an $s$-channel scalar mediator.
The most simplistic version of this scenario involves the introduction
of a single new scalar, $S$, which mediates interactions between
$\overline\chi \chi$ and $\overline{f} f$.  However, because we assume
the DM is not charged under the SM gauge symmetries, while
$\overline{f} f = \overline{f}_L f_R + h.c.$ transforms as a $SU(2)_L$
doublet, this setup is not gauge invariant.  This problem can be
solved by mixing of the singlet scalar with the SM Higgs doublet, as
in Refs.\citep{Khoze:2015sra,Baek:2015lna,Bauer:2016gys,Robens:2016xkb,Wang:2015cda,Lopez-Val:2014jva,Costa:2015llh,Dupuis:2016fda,Balazs:2016tbi}. Like the
$Z'$ mediator case discussed above, we are thus forced to consider a
two mediator model.  However, as one of the two mixed scalar mediators
is actually the SM Higgs boson, the parameter space of this model is
quite constrained by measurement of the Higgs properties, and because
all couplings to SM fermions are restricted to be proportional to SM
Yukawa couplings.

This motivates a scenario in which the singlet scalar mixes not with
the SM Higgs doublet, but with an {\it additional} doublet in a two
Higgs doublet model (2HDM).  The extension to 2HDM plus singlet has
the advantage that we can approximately decouple the SM Higgs from the
two mixed scalars that participate in DM interactions.  Therefore,
compared with the SM Higgs + singlet case:
\begin{itemize}
\item 
a greater range of scalar masses is possible, as neither of the two
mixed scalars must have the 125 GeV mass of the SM Higgs
\item
the couplings of the additional scalars to the SM fermions is not
necessarily dictated by the SM Yukawa couplings.
\end{itemize}
In particular, we shall explore the scope for having different
proportionality constants for the up-type quarks, down-type quarks and
leptons, or flavour dependent couplings that are not proportional to
the Standard Model Yukawa couplings.

Note that 2HDMs are very well motivated, arising in SUSY, GUTs and
other extensions of the SM, and have been extensively studied
\citep{Branco:2011iw,Amaldi:1991cn,Carena:1995wu,Bhattacharyya:2015nca}.
The inclusion of an extra singlet scalar, 2HDM+S, occurs in the NMSSM
version of SUSY \citep{Fayet:1976et}, where in that application the
singlet scalar is added to solve the $\mu$ problem
\citep{Gunion:1984yn,King:2012tr}.  While the 2HDM+S scenario
\citep{Chen:2013jvg,Kanemura:2015fra,vonBuddenbrock:2016rmr} will allow a broader range of
phenomenological possibilities for our dark matter application, it
will also be subject to constraints.  Chief among these, as with all
2HDMs, will be flavour changing bounds. We must therefore choose
particular schemes in which potentially dangerous flavour changing
couplings are kept under control, as will be discussed in detail
below.

We shall explore the phenomenology of these mixed scalar mediator
models to determine viable regions of parameter space that are
consistent with current direct detection bounds.  One feature of such
models will be the interference of the amplitudes governed by the two
mediators, which in some cases will lead to relative cancellation and
hence suppression of cross sections.  Such destructive interference is
a generic feature of multi-mediator models, and has been discussed in
Refs.~\citep{Bauer:2016gys,Ko:2016ybp,Duerr:2016tmh}, for the SM Higgs
+ singlet case.  In our case, the effect is expanded to cover scalar
masses that are unrelated to that of the SM Higgs.
In addition, relative cancellations of contributions from
different quarks will also suppress direct detection cross sections in
some of the scenarios we consider.

The parameter space which survives the direct detection bounds will be
of interest in the upcoming LHC run II analyses.  Note that it is
non-trivial to recast existing mono-X bounds to apply to the mixed
scalar, due to the presence of the two interfering mediators (the
exception, of course, is where one of the scalar mediators can be
taken to decouple, effectively leaving a single mediator) and also due to the presence of additional signals, such as W/Z + MET and VBF + MET, which would
contribute to the jets + MET signal. We defer a
detailed collider analysis of the mixed scalar scenario to a future
publication. 

For scalar mediator models, indirect detection signals are expected to
be too small to provide interesting constraints, as all the DM
annihilation modes are p-wave suppressed by a factor of $v^2 \sim
10^{-6}$ in the present universe (The p-wave suppression applies to
both the $\overline\chi \chi \rightarrow \overline{f} f $ annihilation
via a scalar mediator, and also the $\overline\chi\chi \rightarrow SS$
annihilation to a pair of scalars \citep{Kumar:2013iva}).  Note, though, that p-wave
processes may play a role at freezeout where the $v^2$ suppression
factor is much less severe.  Even so, it will be difficult to obtain
the correct relic density in the SM Higgs + singlet model, except
perhaps on resonance.  The less restrictive 2HDM+S model allows more
freedom to reproduce the correct relic density, either via an
s-channel resonance in $\overline\chi \chi \rightarrow \overline{f} f$
(i.e. $2m_\chi \simeq m_{S}$, where $m_S$ is the mass of one of the
mixed scalar states) or via annihilation into the scalars.  Of course,
adding additional particles into which the DM can efficiently
annihilate, and which ultimately decay to the SM, is a way one can
always fix the relic density, albeit at the expense of simplicity.

The case of a pseudoscalar mediator is very closely related to the
scalar mediator models we consider.  As with the scalar, the
interaction of gauge singlet $\overline{\chi}\chi$ with an
$\overline{f} \gamma_5f$ bilinear can only occur via the mixing of a
SM-singlet pseudoscalar with an $SU(2)_\mathsc{L}$ doublet.  However,
given that the SM contains no pseudoscalar (after electroweak symmetry
breaking), a mixed pseudoscalar plus 2HDM is the minimal scenario.
Such models are discussed in detail in
\citep{Ipek:2014gua,Berlin:2015wwa,Goncalves:2016iyg,No:2015xqa,Bauer:2017ota,Haisch:2016gry}.

For the sake of clarity we shall use the following naming scheme to
refer to the various scalar mediator models:
\begin{itemize} 
\item

Singlet scalar model (S).  This is the non gauge invariant, single
mediator, Simplified Model.
\item 
Singlet scalar plus SM Higgs mixing model (H+S).
\item
Singlet scalar plus two Higgs doublet model (2HDM+S).
\end{itemize} 
The outline of the paper is as follows: We discuss the S and H+S
models and their constraints in Section~\ref{sec:gauge}, and outline
possible 2HDM+S extensions to these scenarios in
Section~\ref{sec:2higgs}.  Direct detection bounds on all models are
presented in Section~\ref{sec:direct}.

\section{Restoring Gauge Invariance in S-channel Simplified Models}
\label{sec:gauge}

The simplified models used by ATLAS and CMS for Dark Matter Searches
(and included in recommendations for Run 2 \citep{Boveia:2016mrp}) are
usually defined after electro-weak symmetry breaking. The standard
framework for imposing MFV is to require that all couplings to quarks
are proportional to the Standard Model Yukawas couplings,
$y_i$~\citep{Abdallah:2014hon,Abdallah:2015ter,Abercrombie:2015wmb,Boveia:2016mrp}. The
resulting Lagrangian is
\begin{eqnarray} 
\label{eq:scalarSM}
L &=& L_{\mathsc{sm}} + \frac{1}{2} \partial^\mu S \partial_\mu S -
\frac{1}{2} M^2 S^2 -V_{\text{int}}(S) -g_q S\sum_q
\frac{y_i}{\sqrt{2}} \bar{q}_i q_i +
\bar{\chi}(i \fsl{\partial}-\tilde{m}_\chi)\chi -y_{\chi}S\bar{\chi}\chi 
\end{eqnarray}
where $\chi$ is the DM, $q_i$ are the SM quarks, and $S$ is a singlet
scalar mediator.  The interaction part of the potential,
$V_{\text{int}}(S)$, is usually neglected.

This Lagrangian is not invariant with respect to the $SU(2)_L\times
U(1)_Y$ gauge symmetries.  If $\chi$ is a SM singlet, the
$S\bar{\chi}\chi$ vertex requires $S$ to also be a singlet.  However,
the SM bilinears $\bar{Q}_Lu_R$ and $\bar{Q}_Ld_R$ are not SM
singlets, and can couple only to scalars that have the same quantum
numbers as a Higgs doublet.  Gauge invariance can be restored by
allowing the singlet scalar $S$ to mix with either the SM Higgs (H+S
model), or with an additional Higgs doublet (2HMD+S model).  We shall
outline the H+S model, and its constraints, and then turn to
investigate whether it is possible to relax those constraints in the
more general 2HDM+S framework.

In the H+S model, the scalar potential before EW symmetry breaking is given by 
\begin{equation} 
V = - \frac{1}{2} M_{SS}^2 S^2 + \mu_{HS} \Phi^\dagger\Phi S +
\frac{1}{2} \lambda_{HS}\Phi^\dagger\Phi S^2 + \frac{1}{3!} \mu_S S^3
+ \frac{1}{4!} \lambda_S S^4.  
\end{equation}
This potential is stable for
$\lambda_{HS} > - \sqrt{\frac{2}{3}\lambda \lambda_S}$, 
where $\lambda$ is the SM Higgs quartic coupling. One may impose a
$Z_2$ symmetry on the potential for $S$, thus discarding the $S^3$ and
$\Phi^\dagger\Phi S$ terms.  Although we shall make this choice for
simplicity, we note that including these additional parameters could
allow more freedom to enhance certain signals.

To permit the new scalar to couple to quarks, it needs to mix with the
Higgs after EWSB, which requires that $\lambda_{HS}$, $\langle \phi
\rangle=v$ and $\langle S \rangle=w$ are nonzero.  The condition
$M_{SS}^2 >\frac{1}{2}\lambda_{SH} v^2$ guarantees that $S$ acquires a vev.  Defining the fields after symmetry breaking as
\begin{equation}
  \Phi = \left(
\begin{array}{cc}
 G^+ \\
\frac{v + h' + G^0}{\sqrt{2}} \\
\end{array}
\right) \,\,\,\, \textrm{and} \,\,\,\,\,
S = w + s',
\end{equation}
the mass matrix becomes
\begin{eqnarray}
\label{HPmassmatrix} M^2 = \left(
\begin{array}{cc}
 2 \lambda v^2   & \lambda_{SH} v w  \\
 \lambda_{SH} v w  & \frac{1}{3} \lambda_{S} w^2\\
\end{array}
\right).   
\end{eqnarray}
This mass matrix can be diagonalised via a rotation to the mass
eigenstate fields, $h= \cos\epsilon h' + \sin\epsilon s' $ and
$s=-\sin\epsilon h' + \cos\epsilon s'$.  As the Higgs Boson observed
at the LHC is very SM-like, we take the mixing angle $\epsilon$ to be small by
considering only small values for $\lambda_{SH}$.  In this limit the
mass eigenvalues are 
\begin{equation}
M_h^2 = 2 \lambda v^2  + O(\lambda_{SH}^2) 
 \,\,\,\, \textrm{and} \,\,\,\,\,
M_S^2 = \frac{1}{3}\lambda_S w^2 + O(\lambda_{SH}^2),
\end{equation}
with mixing angle 
\begin{equation} 
\tan\epsilon \simeq \frac{3 \lambda_{SH} v w}{6 \lambda
  v^2-\lambda_{S} w^2} \simeq \frac{\lambda_{SH} v w}{M_h^2-M_S^2}
\simeq \sin\epsilon .\label{eq:mixepsilon}
\end{equation}
The $h'-s'$ mixing allows both the $h$ and $s$ couple to Standard
Model fermions and dark matter
\begin{equation}
L_{\text{int},\text{$s$-$h$}}=- h \cos\epsilon  \sum_q \frac{m_i}{v} \bar{q}_i q_i + s \sin\epsilon \sum_q \frac{m_i}{v} \bar{q}_i q_i  - y_{DM}(s \cos\epsilon  + h \sin\epsilon ) \bar{\chi}\chi .
\label{eq:int}
\end{equation}
The quark-$s$ couplings coincide with the ones of eq.~(\eqref{eq:scalarSM}) with 
\begin{equation}
g_q \equiv -\sin\epsilon \label{eq:gqtosin} .
\end{equation}
In this model, the proportionality constant $g_q$ is universal across
generations.  Indeed, the same proportionality constant applies not
only for up and down quarks, but also for leptons. Importantly, both
$s$ and $h$ mediate interactions between quarks and DM, so we have a
two-mediator model.

When $m_\chi < M_h/2$, the decay of the SM Higgs to DM gives a
contribution to the Higgs invisible width of
\begin{equation}
\Gamma_{h\rightarrow\chi\bar{\chi}} = \frac{y_\chi^2
  \sin^2\epsilon}{8\pi} M_h \left(1-\frac{4m_\chi^2}{M_H^2}
\right)^{3/2}, 
\end{equation} 
which places tight constraints in this region of parameter space.
ATLAS and CMS report upper limits of $0.78$ \citep{Aad:2015uga} and
$0.58$ \citep{Chatrchyan:2014tja} at 95\% C.L. on the invisible
branching fraction of the Higgs-boson. A combined analysis
\citep{ATLAS-CONF-2015-044} also reports a lower bound on the Higgs
signal strength $\mu$ of $0.87$.  The quantity $\sin\epsilon$ is
currently constrained by the Higgs signals strength to be
$\sin\epsilon < 0.4$ when $2m_\chi < M_h$; for $2m_\chi > M_h$ the
Higgs invisible width provides no useful constraint.
In addition to the interaction terms of eq.~(\ref{eq:int}), there are
other terms that are first order in $\sin\epsilon$, such as
\begin{equation}
L_{\text{int},swz}=-\sin\epsilon \left(2\frac{M_W^2}{v}W_\mu^+ W^{-\mu} +\frac{M_z^2}{v} Z_\mu Z^\mu \right)s 
\label{eq:vecbosfusop}.
\end{equation}
All Standard model processes are only affected at second order in
$\epsilon$ (apart from the ones involving Higgs cubic and quartic
couplings).  Nonetheless, this model is highly constrained by SM
physics. For instance, precision electroweak constraints place upper
limits on the mixing angle that range from $|\sin\epsilon| < 0.4$ when
$M_S \sim 200$~GeV, to $|\sin\epsilon | < 0.2$ when $M_S \sim
1000$~GeV~\citep{Lopez-Val:2014jva}.

\begin{figure}[t]
\captionsetup[subfigure]{position=b}
\centering
\subcaptionbox{\label{fig:feynmonojetA}}{\includegraphics[width=0.19\columnwidth]{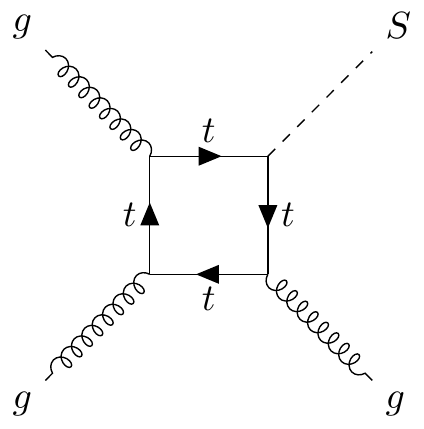} \hspace{1em}}
\subcaptionbox{\label{fig:feynmonojetB}}{\includegraphics[width=0.18\columnwidth]{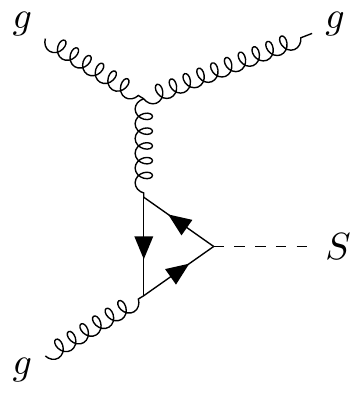} \hspace{1em}}
\subcaptionbox{\label{fig:feynmonojetC}}{\includegraphics[width=0.17\columnwidth]{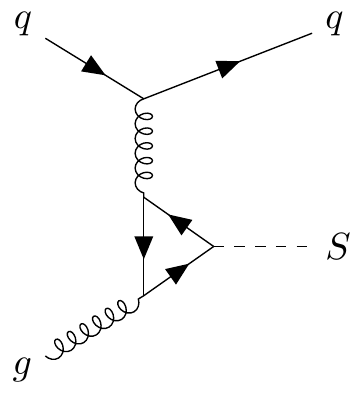} \hspace{1em}} 
\subcaptionbox{\label{fig:feynmonojetD}}{\includegraphics[width=0.20\columnwidth]{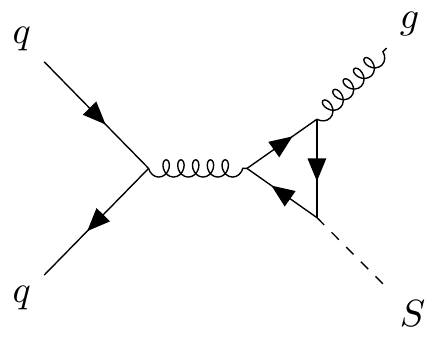} \vspace{1em} \hspace{1em}}\\
\subcaptionbox{\label{fig:feynmonojetE}}{\includegraphics[width=0.15\columnwidth]{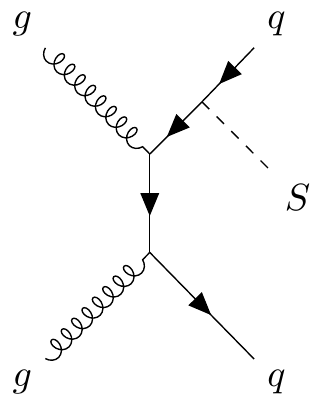}} 
\subcaptionbox{\label{fig:feynmonojetF}}{\includegraphics[width=0.22\columnwidth]{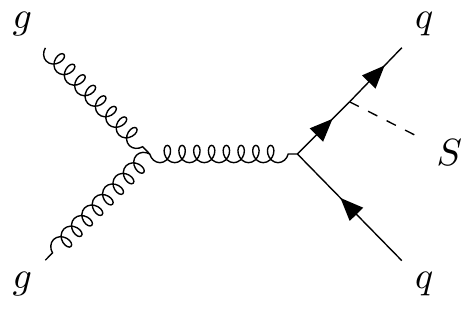} \vspace{1.57em} \hspace{1em}} 
\subcaptionbox{\label{fig:feynmonojetG}}{\includegraphics[width=0.15\columnwidth]{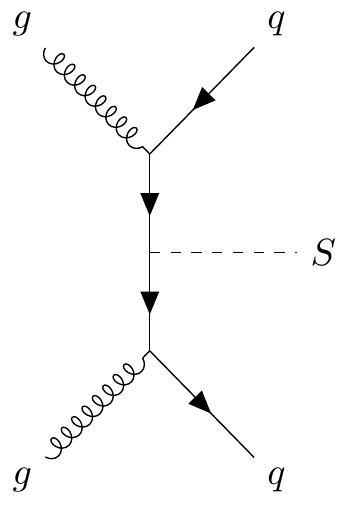}\hspace{1em}}
\subcaptionbox{\label{fig:feynmonojetH}}{\includegraphics[width=0.15\columnwidth]{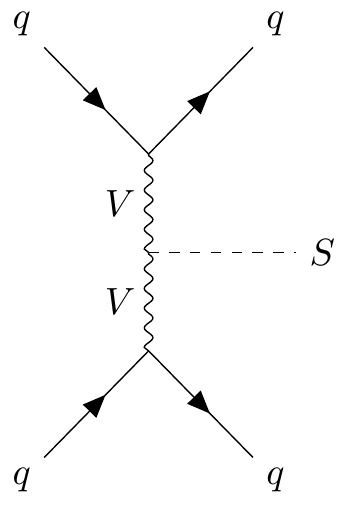}} 
\caption{Feynman diagrams contributing to jets/mono-jet$+\met$ signals
  in the S+H model.  Diagrams
  (\subref{fig:feynmonojetE})--(\subref{fig:feynmonojetH}) also
  contribue to $t\bar{t}+\met$.  All diagrams except
  (\subref{fig:feynmonojetH}) are also present in the S model.}
    \label{fig:feynmonojet}
\end{figure}

\begin{figure}[t]
\captionsetup[subfigure]{position=b}
\centering
\subcaptionbox{\label{fig:feynmonoWZHA}}{\includegraphics[width=0.21\columnwidth]{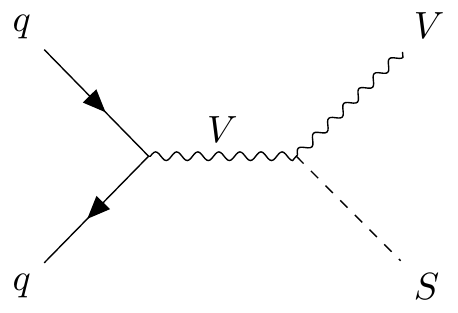} \hspace{1em}}
\subcaptionbox{\label{fig:feynmonoWZHB}}{\includegraphics[width=0.19\columnwidth]{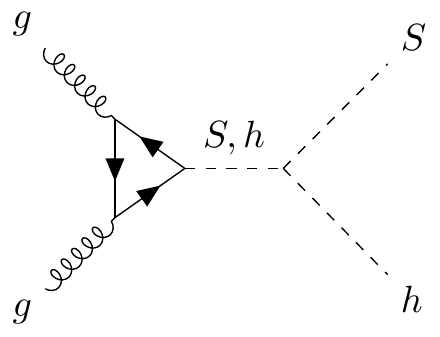} \hspace{1em}}
\subcaptionbox{\label{fig:feynmonoWZHC}}{\includegraphics[width=0.19\columnwidth]{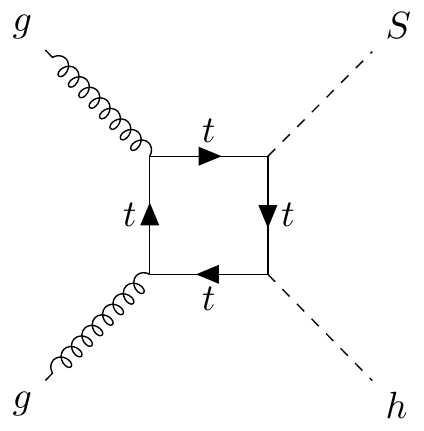} \hspace{1em}} 
\caption{Feynman diagrams for mono-$W/Z+\met$ (\subref{fig:feynmonoWZHA}) and mono-Higgs (\subref{fig:feynmonoWZHB} and \subref{fig:feynmonoWZHC}).}
    \label{fig:feynmonohw}
\end{figure}

Benchmark studies for this model can be found
in~\citep{Robens:2016xkb,Bauer:2016gys}, and relevant diagrams for
monojet$+\met$ and $t\bar{t}+\met$ are shown in
Fig.~\ref{fig:feynmonojet}.  In addition, this model can lead to
mono-$W/Z+\met$, $VBF+\met$ and even mono-Higgs$+\met$ processes,
provided that we keep all the new terms arising from the scalar
potential. Mono-Higgs processes depend on the trilinear vertices $hhS$
and $hSS$, and the relevant Feynman diagram is shown in
Fig. \ref{fig:feynmonohw}.

The Monojet and $t\bar{t}+\met$ sensitivity for the H+S model can be
roughly estimated by rescaling that for the singlet S model, using
\begin{equation}
\frac{\sigma_{S+H}}{\sigma_{S}} \sim \frac{\sin^2\epsilon}{g_q^2} \lesssim \frac{0.4^2}{g_q^2}. 
\label{eq:rescale}
\end{equation} 
Note that such a rescaling assumes the cross section are dominated by
the new scalar mediator.  If both the $s$ and $h$ mediators
contribute, interference effects make such a rescaling invalid. Also
note that this does not account for the contribution of the vector
boson fusion operators of eq.~(\ref{eq:vecbosfusop}) to the cross section in
the H+S model, so this approximation is valid as long as such
operators are subdominant.
The singlet S model has a rather small cross section, therefore the
current exclusion limits are weak.  CMS limits derived from
$b\bar{b}+\met, t\bar{t}+\met$ and $jV+\met, V\rightarrow q\bar{q}$
final states are reported in
\citep{CMS:2016uxr,CMS:2016pod,CMS:2016mxc} while ATLAS limits can be
found in \citep{ATLAS-CONF-2016-086,ATLAS-CONF-2016-077}.
CMS has nearly reached the sensitivity to exclude low mass mediators, for $g_q^{sens} \sim 1$.
Using eq.~(\ref{eq:rescale}), we estimate that the cross sections
for the H+S model will be smaller by a factor of
$\sin^2\epsilon/(g_q^{sens})^2 \lesssim 1/6.25$.  This will make
the model much harder to exclude, as it would require a luminosity roughly
$6.25^2 \sim 39$ times larger than for the singlet model,
assuming a statistically limited scenario.

\section{Going beyond with 2HDM(+S)}
\label{sec:2higgs}

We now explain how to go beyond the standard set-up for section 2, in
a scenario where gauge invariance is retained, but restrictions
imposed by SM Higgs properties are relaxed or removed.  In particular
we shall explore the scope to:
\begin{itemize}
\item
Have greater freedom for the couplings of the scalar mediator(s) to the SM fermions, including different
proportionality constants for the coupling to the up quark, down quark and leptons sectors, or 
flavour-dependent couplings not forced to be proportional to SM Yukawa couplings.
\item Have a range of scalar masses unconnected to the SM Higgs mass.
\end{itemize}
To achieve these aims, additional freedom with both the scalar mixing
and Yukawa couplings is required.  This can be achieved by adding a
second Higgs doublet, thereby expanding the scalar sector to that of a
2HDM~\citep{Branco:2011iw} plus a singlet scalar~\footnote{Note that
  such a Higgs sector also arises in the NMSSM, where an additional
  singlet is added to solve the $\mu$ problem. In the NMSSM, however,
  the 2HDM couplings are forced to be of Type-II, while we will
  consider a broader and less restrictive range of Yukawa
  structures.}.  Below, we shall analyse the scalar spectrum of
this scenario, and outline possible Yukawa structures that are
consistent with flavour constraints.  In doing so, we shall need to
review and expand the pertinent features of 2HDMs; Readers who are
familiar with 2HDMs may wish to skip to the direct detection
analysis in section~\ref{sec:direct}.

\subsection{2HDM+S Scalar Spectrum}
\label{sec:2hdmmassspectrum}
The most general scalar potential we consider is
\begin{equation}
V(\Phi_1,\Phi_2,S) = V_{\mathsc{2hdm}}(\Phi_1,\Phi_2) + V_S(S) + V_{S\mathsc{2hdm}}(\Phi_1,\Phi_2,S), 
\end{equation}
where\footnote{Note that there are different conventions for the normalization of these coefficients; we have chosen the convention assumed in \cite{Branco:2011iw}.}\nobreak
\bea
V_{\mathsc{2hdm}}(\Phi_1,\Phi_2) &=& M_{11}^2 \Phi_1^\dagger \Phi_1 + M_{22}^2 \Phi_2^\dagger \Phi_2 +  (M_{12}^2 \Phi_2^\dagger \Phi_1 + h.c.) + \frac{\lambda_1}{2} (\Phi_1^\dagger \Phi_1)^2 + \frac{\lambda_2}{2} (\Phi_2^\dagger \Phi_2)^2 
\nonumber \\
&+&\lambda_3 (\Phi_1^\dagger \Phi_1)(\Phi_2^\dagger \Phi_2) + \lambda_4 (\Phi_2^\dagger \Phi_1)(\Phi_1^\dagger \Phi_2)  
\nonumber \\
&+& \frac{1}{2}\left(\lambda_5 (\Phi_2^\dagger \Phi_1)^2  + \lambda_6(\Phi_2^\dagger \Phi_1) (\Phi_1^\dagger \Phi_1) +  \lambda_7(\Phi_2^\dagger \Phi_1)(\Phi_2^\dagger \Phi_2) + h.c.\right), 
\\
V_S(S) &=& \frac{1}{2} M_{SS}^2 S^2 + \frac{1}{3} \mu_S S^3 + \frac{1}{4} \lambda_S S^4,
\\
V_{S\mathsc{2hdm}}(\Phi_1,\Phi_2,S) &=& \mu_{11S}(\Phi_1^\dagger \Phi_1)S + \mu_{22S}(\Phi_2^\dagger \Phi_2)S + (\mu_{12S} \Phi_2^\dagger \Phi_1 S + h.c.) 
\nonumber\\
&+& 
\frac{\lambda_{11S}}{2}(\Phi_1^\dagger \Phi_1)S^2 +  \frac{\lambda_{22S}}{2}(\Phi_2^\dagger \Phi_2)S^2 + \frac{1}{2}(\lambda_{12S} \Phi_2^\dagger \Phi_1 S^2 + h.c.).
\eea

To reduce the complexity of this model, we make several simplifying
assumptions which are common with those made in many 2HDM studies.  We
assume CP is conserved in the Higgs sector, and is not spontaneously
broken by a relative phase between the vevs of the 2 doublets.  This
implies that $M_{12},\lambda_5,\mu_{12S}, \lambda_{12S}$ are real.
We also impose a $Z_2$ symmetry on the potential $V_{\mathsc{2hdm}}$,
under which one of $\Phi_{1,2}$ is odd while the other is even,
eliminating the $\lambda_{6,7}$ terms.  However, we allow the $Z_2$
symmetry to be broken by soft terms, and thus keep the $M_{12}$ term.
Moreover, as we are interested in building up a simplified model, we
are primarily interested in the scalar mass spectrum and mixing, but
not the scalar interactions.  From this point of view, the parameters
$\lambda_{ijs}$ and $\mu_{ijs}$ are equivalent~\footnote{This is true
  for models with scalar mixing, but not for a pseudoscalar model, as the pseudoscalar gets no vev.} so we choose to set the latter to
zero. The same is true for $M_{SS}$ and $\mu_{S}$, so we discard the latter.

By writing the fields as
\bea
\Phi_i = \left(
\begin{array}{cc}
 \Phi_i^+ \\
\frac{v_i + \rho_i + i \eta_i}{\sqrt{2}} \\
\end{array}
\right), \,\,\,\,\,\,\,\, S = v_S + \rho_3, \eea and using the mimima
condition for the potential, we can eliminate the parameters $M_{11}$,
$M_{12}$, $M_{SS}$ and replace them with $v_1$, $v_2$, $v_S$.  The
mass matrix for the scalars is built by a 2x2 block for the charged
scalars (one of which will be the SM Goldstone boson $G^+$ for the
$W$), a 2x2 block for the pseudoscalars (one of which will be the SM
Goldstone boson $G^0$ for the $Z$) and a 3x3 block for the 3 scalars.
Defining the ratio of vevs in the usual way, 
\begin{equation}
\tan\beta =  \frac{v_2}{v_1}, \;\;\;\; {\textrm{with}} \;\;\;\; 
v_1^2+v_2^2 = v^2,
\end{equation}
the non-zero eigenvalues for the physical charged scalar and pseudoscalar are 
\begin{eqnarray}
M_{H^+}^2 &=& \sec ^2\beta \left(M_{22}^2+\frac{1}{2} \lambda_{22S} v_S^2\right)+\frac{v^2}{2} \left(\lambda_2 \tan ^2\beta+\lambda_{3}\right),\\
M_A^2 &=& \sec ^2\beta \left(M_{22}^2+\frac{1}{2} \lambda_{22S} v_S^2\right)+ \frac{v^2}{2} \left(\lambda_2 \tan ^2\beta+ 
   \lambda_3+\lambda_4-\lambda_5 \right).
\end{eqnarray}
It is useful to perform a rotation to the so called ``Higgs basis" where only one doublet obtains a vev
\bea
\Phi_h = \cos\beta \Phi_1 + \sin\beta \Phi_2 = \left(
\begin{array}{cc}
 G^+ \\
\frac{v + h + i G^0}{\sqrt{2}} \\
\end{array}
\right),\label{eq:alignh}\\
\Phi_H = -\sin\beta \Phi_1 + \cos\beta \Phi_2 = \left(
\begin{array}{cc}
 H^+ \\
\frac{ H + i A}{\sqrt{2}} \\
\end{array}
\right).\label{eq:alignH}
\eea
The mass matrix for the scalars in the $\{h, H, S\}$ basis is 
\bea
M^\rho &=&  \left(
\begin{array}{ccc}
M^\rho_{hh} & M^\rho_{hH} & M^\rho_{hS}\\
 M^\rho_{hH} & M^\rho_{HH} &  M^\rho_{HS}\\
 M^\rho_{hS} & M^\rho_{HS} & M^\rho_{SS}\\
\end{array}
\right),
\label{eq:scalarmass}
\eea
where
\begin{align} 
M^\rho_{hh} &= \frac{\lambda_1+\lambda_2}{2} v^2 +\frac{\lambda _1-\lambda_2}{2}v^2\cos 2\beta -\frac{\lambda_1+\lambda_2-2\lambda_{345}}{4}v^2\sin^2 2\beta, \\
M^\rho_{hH} &= -\frac{1}{4} v^2 \sin2\beta \left(\lambda _1-\lambda _2+\left(\lambda_1+\lambda_2-2\lambda_{345}\right) \cos 2
   \beta\right),\\
M^\rho_{hS} &= \frac{v v_S}{2} \left(\lambda_{11S}+\lambda_{22S}+\cos 2 \beta\left(\lambda_{11S}-\lambda_{22S}\right)+2\lambda_{12S}\sin2\beta \right),\\
M^\rho_{HH} &= \sec ^2\beta  \left(M_{22}^2 +\frac{\lambda_2}{2} v^2 +\frac{\lambda_{22S}}{2} v_S^2\right) + \frac{\lambda_1-\lambda_2}{4}v^2 - \frac{\cos^2 2\beta}{4}\left(\lambda_1+\lambda_2-2\lambda _{345}\right) v^2, \\
M^\rho_{HS} &= \frac{v v_S}{2} \left(\sin 2 \beta\left(\lambda_{22S}-\lambda_{11S}\right)+2\lambda_{12S}\cos 2\beta \right),\\
M^\rho_{SS} &= 2\lambda_S v_S^2,
\end{align}
and we have defined $\lambda _{345}=\lambda_3+\lambda_4+\lambda_5$.

We shall now impose the alignment limit \citep{Gunion:2002zf,Carena:2013ooa} in the context of this
model, as this is the experimentally favoured configuration in which
one of the scalars couples like the SM Higgs boson.  In a standard
2HDM, the definition of alignment limit is that $M^\rho_{hH} = 0$ so
that both $h$ and $H$ are mass eigenstates\footnote{In 2HDM, the
  alignment limit can arise naturally in the presence of a softly
  broken CP2
  symmetry~\citep{Dev:2014yca,Dev:2015bta,Pilaftsis:2016erj,Draper:2016cag}
  that imposes $\lambda_1=\lambda_2=\frac{1}{2}\lambda
  _{345}$ (or $\lambda_1=\lambda_2=\lambda_{345}$ for the normalization used in this paper). However, in the presence of the additional singlet we are
  not aware of the existence of any symmetry that can make this
  alignment to arise naturally. Nonetheless, we enforce it by
  requiring the non-diagonal entries of the first row/column of
  Eq.~(\ref{eq:scalarmass}) to vanish.}.  In the case of the 2HDM+S, the
presence of 3 mixed scalars makes the alignment limit more
complicated.  We shall enforce alignment by requiring that
$M^\rho_{hH} = M^\rho_{hs} = 0$, which implies
\begin{align} 
\lambda_3 &=
  \frac{1}{2}\left(\lambda_1+\lambda_2-2\lambda_4-2\lambda_5
  +\left(\lambda_1-\lambda_2\right)\sec
  2\beta\right),
\label{eq:alignment}\\ 
\lambda_{11S} &=
  -\tan\beta\left(2\lambda_{12 S}+\lambda_{22S} \tan\beta\right),
\end{align}  
and sets the $h$ in Eq.~(\ref{eq:alignh}) to be the mass eigenstate
which corresponds to the SM Higgs, with $M_h = M_{hh}^\rho$.  While
some deviation from exact alignment would be compatible with current
Higgs measurements, we adopt this approximation for simplicity.
The remaining 2x2 matrix can be diagonalized to obtain mass
eigenstates $\{S_1,S_2\}$, with eigenvalues and mixing angle given by
\begin{align}
M_{S_{1,2}}^2 &=
\frac{1}{2}\left(M_A^2+\lambda_5 v^2+ \left(\lambda_2
v^2-M_h^2\right)\tan^2 \beta\right)\left(1\pm \frac{1}{\cos
  2\theta}\right) + \lambda_S v_S^2\left(1\mp \frac{1}{\cos
  2\theta}\right), \\
\tan2\theta &= \frac{4 v \cos
  ^2\beta v_S \left(\tan \beta \lambda_{22S}+\lambda _{12
    S}\right)}{\cos 2 \beta \left(M_A^2+M_h^2-2 \lambda _S
  v_S^2-\lambda _2 v^2+\lambda_5 v^2\right)+M_A^2-M_h^2-2 \lambda _S
  v_S^2+\lambda _2 v^2+\lambda _5 v^2}. 
\end{align}
We can now rewrite the Lagrangian in terms of the mass eigenstates using Eqs.~\ref{eq:alignh},\ref{eq:alignH} and  
\bea
H &=& \cos\theta S_1 - \sin\theta S_2,\\
S &=& v_S + \sin\theta S_1 + \cos\theta S_2. 
\eea

\subsection{2HDM+S Lagrangian}
\label{sec:2hdms-sm}

The 2HDM+S scenario is described by the following Lagrangian
\bea
L&=&L_{\text{\sc sm}} + L_{DM} + L_{S} + L_{A,H^+},\\
L_{DM} &=& i \bar{\chi}\slashed{\partial}\chi - m_\chi \bar{\chi}\chi -y_\chi\left(\sin\theta S_1 + \cos\theta S_2\right) \bar{\chi}\chi,\\
L_s &=&\sum \frac{1}{2} \partial_\mu S_i \partial^\mu S_i - \frac{1}{2} M_i^2 S_i^2 -\sum_{f=u,d,l} \epsilon^f \sum_{i\in f} \frac{y_i}{\sqrt{2}} (\cos\theta S_1 - \sin\theta S_2)\bar{f}_i f_i,\\
L_{A,H^+} &=& \frac{1}{2} \partial_\mu A \partial^\mu A  - \frac{1}{2} M_A^2 A^2 + \partial_\mu H^+ \partial^\mu H^- - M_{H^+}^2 H^+ H^- + i \sum_{f=u,d,l} \epsilon^f \sum_{i\in f} \frac{y_i}{\sqrt{2}} A \bar{f}_i \gamma_5 f_i, \nn\\
&-& \left(\sum_{i\in u, j\in d} \left(y_i V_{ij} \epsilon^u P_L + y_j V_{ij} \epsilon^d P_R\right)\bar{u}_i d_j H^+ +\sum_{i\in l^-} y_i \epsilon^L_i \bar{\nu_i} P_R l_i H^+ + h.c.\right),
\eea
where $V=W,Z$, and we have neglected to write down the 3 and 4 point
scalar interactions, and 4 point $S_i^2 V^2$ interactions.  The
parameters $\epsilon^f$ depend upon the choices made for the Yukawa
couplings to the two doublets, and will be discussed in detail in the
following subsections.  They are given in Table \ref{tab:coeffs} for a
number of common 2HDMs Yukawa structures.

For the purpose of DM phenomenology, the important terms in the
Lagrangian are the coupling of the DM and SM fermions to the two mixed
scalars, $S_{1,2}$. These terms have a similar form to the H+S model
of section~\ref{sec:gauge}, but now neither mediator is the SM Higgs,
and greater coupling freedom has been obtained.  The price we pay for
this additional freedom is the introduction of flavour changing
interactions.  While $L_s$ and $L_{DM}$ are the relevant terms for
direct detection and collider mono-jet searches, the full Lagrangain
should be used if one wishes to impose complementary limits (like
mono-W/Z, mono-Higgs, heavy resonances).  Flavour constraints will
arise primarily from $L_{A,H^+}$.

If there is some hierarchy between the 2 scalar masses (for example,
$M_2 \sim 5 \TeV$, $M_1 \sim 500 \GeV)$, interactions will be
dominated by the exchange of the lighter scalar, while the heavier one
can be approximately decoupled.  In this limit we reproduce the
structure of the singlet scalar model, from a gauge invariant
framework, while retaining additional coupling freedom.  In general,
both scalar mediators will need to be retained.

\subsection{Yukawa structure}
\label{sec:yukawa}

The Yukawa interactions of the SM fermions with the Higgs doublets can
be expressed as
\begin{equation}
L_{\text{Yukawa}} = - \sum_{n=1,2} \left(Y_{n,ij}^U \bar{Q}_L^i u_R^j \widetilde{\Phi}_n
+ Y_{n,ij}^D \bar{Q}_L^i d_R^j \Phi_n
+ Y_{n,ij}^L \bar{L}_L^i l_R^j \Phi_n + h.c. \right).
\label{eq:yukawah1h2}
\end{equation}
As in standard 2HDMs, we shall need to choose Yukawa structures that
keep potentially dangerous flavour violating processes under control.
We outline the possibilities below, and explore the dark matter
phenomenology of these choices in section~\ref{sec:direct} by
determining direct detection constraints.

\subsubsection{Type I,II,X and Y}
\label{sec:type12}

In 2HDMs of type I, II, X and Y, flavour violating processes are
suppressed by imposing a symmetry which permits each type of fermion
(up quarks, down quarks and leptons) to couple to only one of the
Higgs doublets.  This hypothesis is called Natural Flavour
Conservation (NFC), and together with MFV guarantees that FCNCs are
strongly suppressed.  However, the presence of the charged scalar
$H^+$ still allows FCNCs at loop level. This places some constraints
on the parameter $\tan\beta$, that is otherwise experimentally
unconstrained in the Higgs-alignment limit.

In type I, all SM fermions couple to $\Phi_2$, while in type II the up
quarks couple to $\Phi_2$ and the down quarks and leptons couple to
$\Phi_1$.  Type X and Y have the same quarks coupling as type I and II
respectively, but the leptons couple to the opposite doublet.
Therefore, at large $\tan\beta$, Type I and Y will be less constrained
by di-lepton resonance searches than Type X and II (which,
respectively, have the same quark couplings).

\begin{table}[tb]\centering
\begin{tabular}{|c|c|c|c|}
\hline
Model & $\epsilon_d$ & $\epsilon_u$ & $\epsilon_l$
\\ \hline
Type I & $\cot\beta$ & $\cot\beta$ & $\cot\beta$
\\
Type II & $-\tan\beta$ & $\cot\beta$ & $-\tan\beta$
\\
Type X & $\cot\beta$ & $\cot\beta$ & $-\tan\beta$
\\
Type Y & $-\tan\beta$ & $\cot\beta$ & $\cot\beta$
\\
Inert & 0 & 0 & 0
\\ \hline
\end{tabular}
\caption{Values of the coefficients $\epsilon_{u,d,l}$ which
  correspond to models with discrete ${\cal Z}_2$
  symmetries.}\label{tab:coeffs}
\end{table}

\subsubsection{Type III models and Minimal Flavour Violation}
\label{sec:type3}
In type III 2HDMs, there is no symmetry to forbid the fermions from
coupling to both doublets, and so no NFC assumption. It is well known
that in type III 2HDMs, in absence of some additional mechanism, FCNCs
can arise at tree level. To avoid this we will implement Minimal
Flavour Violation (MFV), adopting the most general version of MFV
following~\citep{D'Ambrosio:2002ex,Buras:2010mh}.
For the type III scenario, it makes no sense to use the basis
$\Phi_{1,2}$ so we will instead work in the Higgs basis $\Phi_{h,H}$
where it will be easier to describing the couplings.  Rewriting
Eq.~(\ref{eq:yukawah1h2}) in this basis we have
\begin{equation}
L_{\rm Yukawa} = - \sum_{n=h,H} \left(Y_{n,ij}^U \bar{Q}_L^i u_R^j \widetilde{\Phi}_n
+ Y_{n,ij}^D \bar{Q}_L^i d_R^j \Phi_n
+ Y_{n,ij}^L \bar{L}_L^i l_R^j \Phi_n + h.c. \right),
\label{eq:yukawahH}
\end{equation}
where the matrices $Y_{h,ij}^{U,D,L}$ have to be the SM Yukawa
matrices.  Written in an arbitrary fermion basis, these are arbitrary
3x3 complex matrices that only need to reproduce the right mass
eigenvalues and the CKM matrix. We can choose to work in the basis
where they can be written as 
\begin{align}
Y_h^U &= V^\dagger D(y_u,y_c,y_t) = V^\dagger D_U,  \\
 Y_h^D &= D(y_d,y_s,y_b) = D_D,  \\ 
Y_h^L &=D(y_e,y_\mu,y_\tau) = D_L, 
\end{align}
where $D(x_i,...,x_j)$ indicates a diagonal matrix where $x_i,...,x_j$
are the diagonal elements.  To have minimal flavour violation, we
need to impose \citep{D'Ambrosio:2002ex,Buras:2010mh} 
\begin{align}
Y_H^U = P_U(Y_h^U Y_h^{U\dagger}) Y_h^U, \;\;\; 
\end{align}
and similarly for $Y_D, Y_L$, where $P_{U,D,L}(x)$ are generic
polynomials.  Now because $(Y_h^U Y_h^{U\dagger})^n Y_h^U = V^\dagger
D_U^{2n+1}$ and $(Y_h^D Y_h^{D\dagger})^n Y_h^D = D_D^{2n+1}$, this
results in the following Yukawa structure
\begin{align}
Y_H^U &= V^\dagger P_U(D_U^2)D_U = V^\dagger D(\lambda_u,\lambda_c,\lambda_t),\\
Y_H^D &= P_D(D_D^2) D_D = D(\lambda_d,\lambda_S,\lambda_b),\\
Y_H^L &= P_L(D_L^2) D_L = D(\lambda_e,\lambda_\mu,\lambda_\tau).
\end{align}
This Yukawa structure guarantees that it is possible to simultaneously
diagonalize the masses of the fermions and their Yukawa couplings to
both doublets.  Note that unfortunately this structure is not RGE
invariant. Identifying all the allowed coupling patterns that avoid
FCNCs and are stable under quantum corrections is beyond the scope of
this work.  Here we simply point out that there are possible coupling
patterns beyond the type I and II 2HDMs.  We shall illustrate two
specific type III examples below, the Aligned model and a 2-generation
model.  A futher type III possibility, with approximate alignment, is
discussed in Appendix~\ref{sec:ckmexplanation}.

\subsubsection{Aligned 2HDM+S}
\label{sec:a2hdm}

The Aligned 2HDM~\citep{Pich:2009sp,Tuzon:2010vt} is a type-III 2HDM
that interpolates between Type I and II.  While Type I,II models
impose a symmetry to force the fermions to couple to only to one of
the two doublets, in the Aligned model we simply assume the fermions
couple to only one linear combination of $\Phi_{1,2}$, but that such
linear combination can be different for $u,d,l$\footnote{Note that
  this hypothesis is not stable under quantum corrections, so the
  model will not be NFC compliant, while still MFV. Loop corrections
  have been studied in several works
  \citep{Pich:2010ic,Braeuninger:2010td,Botella:2015yfa} and found not
  to be tightly constraining.}.  We can define the linear combinations
that couple to up, down quarks and leptons $\Phi_{u,d,l}$ and their
orthogonal ones $\Phi_{u,d,l\perp}$ to be
\begin{eqnarray}
\left(
\begin{array}{c}
\Phi_i  \\
\Phi_{i\perp}  \\
\end{array}
\right) = 
\left(
\begin{array}{cc}
\cos\gamma_i  & \sin\gamma_i \\
-\sin\gamma_i & \cos\gamma_i \\
\end{array}
\right)
\left(
\begin{array}{c}
\Phi_h   \\
\Phi_H  \\
\end{array}
\right),
\end{eqnarray}
where $i=u,d,l$ and $0\le \gamma_{u,d,l}<\pi$ are arbitrary angles\footnote{References (e.g. \cite{Jung:2010ik,Enomoto:2015wbn,Tuzon:2010vt}) define the Aligned model in terms of the parameters $\zeta_{u,d,l}$, that can be complex in general. Imposing CP conservation forces them to be real and in such a case they are related to the $\gamma_i$ parameters above by $\zeta_i = \tan (\gamma_i)$}.
By requiring the fermion interactions with the Higgs are the same as
in the SM, we identify the coupling to the 2nd doublet 
\begin{align}
Y_h^U & \equiv Y_{\text{\sc sm}}^U,\\
Y_H^U & = \tan\gamma_u Y_{\text{\sc sm}}^U,
\end{align}
and similarly for $d,l$.  With this identification, the fermion
couplings to the new doublet differ from the SM Yukawa couplings by an overall
scaling factor, $\tan\gamma_i$, where $\gamma_i$ can take different
values for $u,d,l$.  This arrangement clearly satisfies the simultaneously
diagonalizable requirement of type III models that respect MFV.  One
can observe that in principle it is possible for the $b$-quark to
couple more strongly to the $S_{1,2}$ mediators than the $t$-quark (this is true
also for Type II with large $\tan\beta$) which suggests the
possibility of different experimental signatures.  In fact, 2HDM of
type I, II, X and Y can be recovered with particular choices for the
angles $\gamma_i$, as listed in Table~\ref{tab:z2models}.

\begin{table}[tb]\centering
\begin{tabular}{|c|c|c|c|}
\hline
Model & $\gamma_d$ & $\gamma_u$ & $\gamma_l$
\\ \hline
Type I & $\pi/2-\beta$ & $\pi/2-\beta$ & $\pi/2-\beta$
\\
Type II & $-\beta$ & $\pi/2-\beta$ & $-\beta$
\\
Type X & $\pi/2-\beta$ & $\pi/2-\beta$ & $-\beta$
\\
Type Y & $-\beta$ & $\pi/2-\beta$ & $\pi/2-\beta$
\\
Inert & 0 & 0 & 0
\\ \hline
\end{tabular}
\caption{Choices of angles $\gamma_{u,d,l}$ which correspond to models with
discrete ${\cal Z}_2$ symmetries.}\label{tab:z2models}
\end{table}

\subsubsection{Coupling to only the first 2 generations (2gen-2HDM+S)}
\label{sec:2GS2HDM}

Until now, all our models have had couplings to quarks that are
proportional to the SM Yukawa couplings. As a consequence the $t$ and
$b$ quark couplings will be dominant, as all other couplings will be
suppressed by a factor of at least $\frac{m_s}{m_b}$. Thus all
the relevant collider phenomenology will be dominated by initial
states with gluons and maybe $b$ quarks, and only final states with
$g,b$ jets or top quarks. On the other hand, most flavour constraints
that enforce this Yukawa structure are coming from the couplings to
the third generation of quarks. Thus it might be interesting to
understand what happens to flavour constraints when one turns off the
couplings to the third family of quarks. Logically, in such case the
phenomenology will be completely different, as $b,t$ will not be
present in the initial and final states, and also $gg$ initial states
with top loop diagrams will no longer contribute.
Ref.~\citep{Botella:2015yfa} analyses in detail the possible Yukawa
structures that can lead to stable RGE solutions. While they focus on
solutions that satisfy the Yukawa alignment condition $Y_H^i \propto
Y_h^i, i=u,d$, the solution 5 that they find in the appendix is a
viable solution also for $Y_i^u \propto Y_i^d, i=H,h$.  Therefore, a
possible Yukawa structure for the additional doublet that may avoid
flavour constraints is\footnote{Yukawa patterns that can yield similar
  enhancements for the signal of the first 2 generations can also be
  found in
  \citep{Evans:2011wj,Altmannshofer:2016zrn,Arroyo:2013tna,Bishara:2015cha}.}
\begin{align}
Y_H^U &= A V^\dagger P_{12},\\
Y_H^D &= B P_{12},
\end{align}
where $A,B$ are real numbers and
\begin{align}
P_{12} =   \left(\begin{array}{ccc}
1 & 0 & 0 \\
0 & 1 & 0 \\
0 & 0 & 0
\end{array}\right),  \;\;\;\;\
P_3 =   \left(\begin{array}{ccc}
0 & 0 & 0 \\
0 & 0 & 0 \\
0 & 0 & 1
\end{array}\right).
\end{align}
The new doublet therefore couples to the first 2 generations of quarks
only, with equal strength.  We neglect the leptons for convenience, as
they can always be taken to be flavour diagonal (or zero) such that no
FCNCs arise in the lepton sector. This structure has been chosen
because it is orthogonal to the one of the SM Yukawa matrices, that,
in the same basis, are 
\begin{eqnarray} 
Y_h^U &=& y_t V^\dagger P_3 + O\left( \frac{m_c}{m_t} \right),
\\ Y_h^D &=& y_b P_3 + O \left( \frac{m_s}{m_t} \right).
\end{eqnarray}
This is a convenient form\footnote{Note that this structure formally
  respects MFV, even though it requires a fine tuning of
  $\mathcal{O}(\frac{y_u}{y_c}, y_c^2)$.} that strongly suppresses
FCNCs, as they must be proportional either to one of the small
parameters $y_u, y_d, y_s, y_c$ or to at least one of the small CKM
matrix elements, $V_{13}, V_{23},V_{31}, V_{32}$.
We can check that this Yukawa structure doesn't induce large FCNC,
provided $A \lesssim 0.1$ and $B\lesssim 0.01$, by solving the RGE
using the formalism of~\citep{Cvetic:1998uw,Desai:2003xz}. This is
done in Appendix~\ref{sec:rge2gen}. Note the this structure can
therefore be used to obtain couplings for $c,s,d,u$ that,
respectively, are up to $13,18,400,7500$ times larger than in the
SM. Given these enhanced couplings, we therefore expect these models
will be subject to much tighter direct detection constraints, unless
some relative cancellation or interference occurs.

\section{Direct Detection}
\label{sec:direct}

Direct Detection (DD) is an efficient way of probing these
models. Interactions mediated by the exchange of scalar particles
generate Spin-Independent operators (SI) that can currently probe
cross sections of the order of $10^{-45} \cm^2$ for DM particles in
the range $10 \GeV < m_\chi < 1 \TeV$. In particular, DD will
typically be more efficient that collider searches in the high-mass
range, $m_\chi\gtrsim 100\GeV$, where collider production cross
sections are significantly suppressed.  However, DD will usually leave
open a small window at low ($m_\chi \lesssim 10\GeV$) DM masses. In
this window, collider physics usually provides stronger constraints.

The gauge invariant class of models considered in this paper introduce
another parameter region where direct detection searches are blind:
When the masses of the scalar mediators are similar, interference
effects will be very
important\citep{Kim:2008pp,Baek:2011aa,LopezHonorez:2012kv}.  This
feature will be present in all the models we considered (in both the
H+S and 2HMD+S scenarios). 
In fact, for degenerate mass mediators, the DD nucleon operators cancel exactly. Since this cancellation is not accidental, but rather arises from a symmetry, it should be expected to hold at loop level. 1-loop corrections were calculated for this model and were indeed found to have negligible contributions to the tree level results presented in this paper.
In the case of 2HDM+S, there will be an
additional source of interference: the interference between different
quarks present in nucleons. For example, in Type II 2HDMs with Higgs
alignment, the up and down quarks have Yukawa couplings of opposite sign,
which will lead to opposite-sign contributions to the coefficient of
the effective nucleon operator.

\subsection{Direct detection constraints for the H+S model}
\label{sec:directsinglet}

This model will be strongly constrained by DD experiments, as the
$125\GeV$ Higgs is one of the mediators. Unless the SM Higgs and second mediator masses are approximately degenerate, interference will play no role and
the DD cross section will be dominated by the exchange of the SM
Higgs, placing a very strong bound on the value of the mixing
angle $\epsilon$. The only relevant nucleon operator will be

\begin{equation}
O_1^N = \bar{\chi}\chi \bar{N}N \label{eq:nuclop},
\end{equation}
with a coefficient related to those for the quark and gluon operators
\begin{align}
O_1^q &= \bar{\chi}\chi \bar{q}q,\\
O_1^g &= \frac{\alpha_s}{12\pi}G_a^{\mu\nu}G_{\mu\nu}^a\bar{\chi}\chi,
\end{align}
according to 
\begin{equation}
c_N = \sum_{q=u,d,s}\frac{m_N}{m_q} f_{T_q}^N c_q + \frac{2}{27} f_{T_g} \sum_{q=c,b,t} \frac{m_N}{m_q} c_q.
\end{equation}
Starting from the Lagrangian at a high energy, we can evolve down to low
energies, integrating out first $s$ and then $h$, to generate the
following EFT operator for light quarks:
\begin{equation}
O_q = c_q O_1^q = \frac{y_q y_\chi \cos\epsilon\sin\epsilon}{\sqrt{2}}\left(\frac{1}{M_h^2}-\frac{1}{M_S^2}\right)\bar{\chi}\chi \bar{q}q. 
\end{equation}
Heavy quarks instead contribute through the gluon operator: 
\begin{equation}
O_g =  -\frac{c_q}{m_q} O_1^g = -\frac{c_q}{m_q} \frac{\alpha_s}{12\pi} G_a^{\mu\nu}G_{\mu\nu}^a\bar{\chi}\chi.
\end{equation}
The coefficient of the nucleon operator \ref{eq:nuclop} is thus
\begin{equation}
c_N = m_N\frac{y_\chi \cos\epsilon\sin\epsilon}{v}\left(\frac{1}{M_h^2}-\frac{1}{M_S^2}\right) \left(\sum_{q=u,d,s} f_{T_q}^N  + \frac{2}{9} f_{T_g} \right).\label{eq:nucleonSH}
\end{equation}
Note that in this scenario,  interference effects between
different quarks are not possible as all fermion couplings to the new scalar
are proportional to the Yukawa couplings to the SM Higgs (with the
same proportionality coefficient)\footnote{This will be true also for
  Type I 2HDM+S, where the same situation applies.}.

 \begin{figure}[!t]
    \centering
    \includegraphics[width=0.45\textwidth]{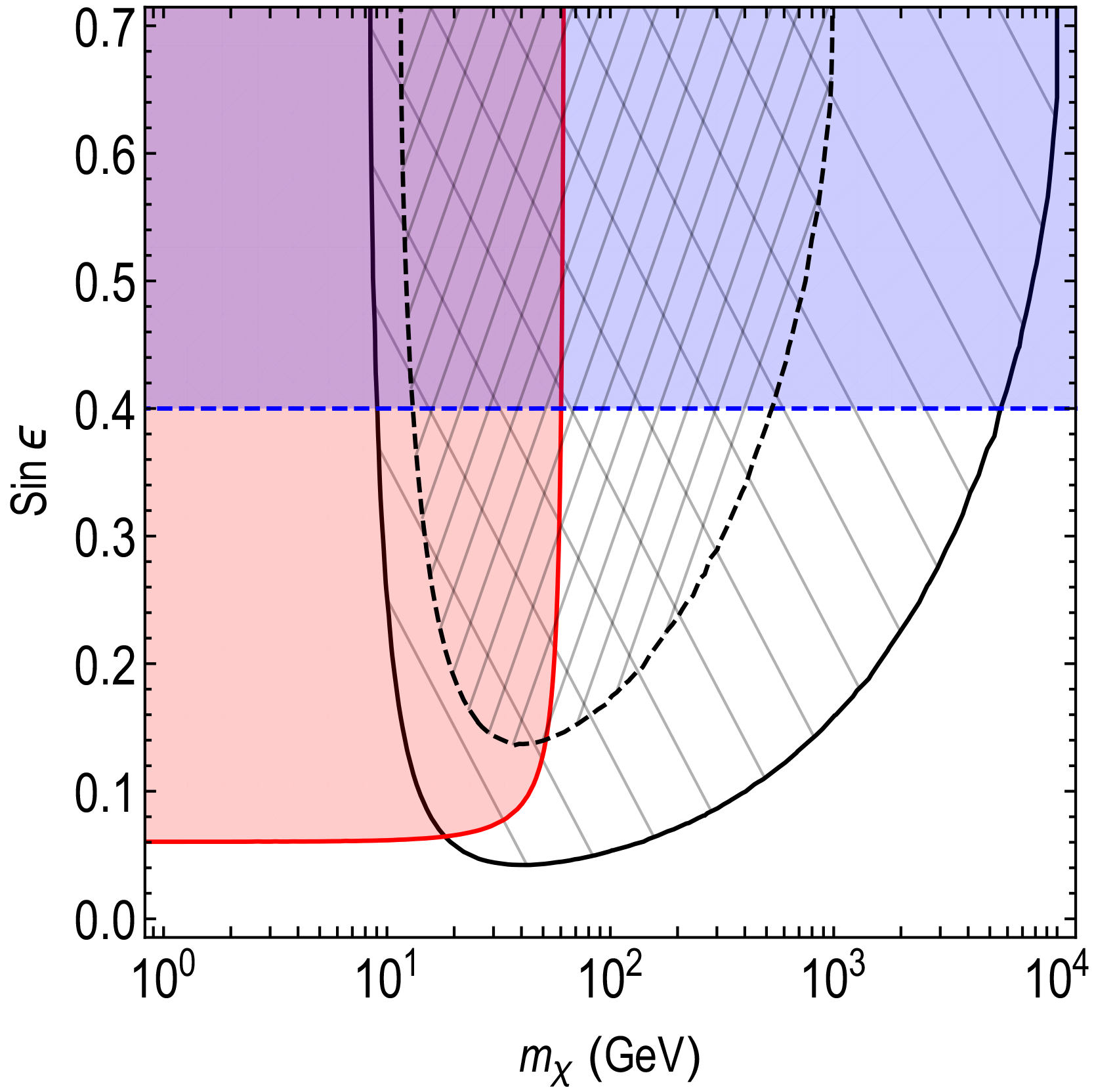}
    \includegraphics[width=0.45\textwidth]{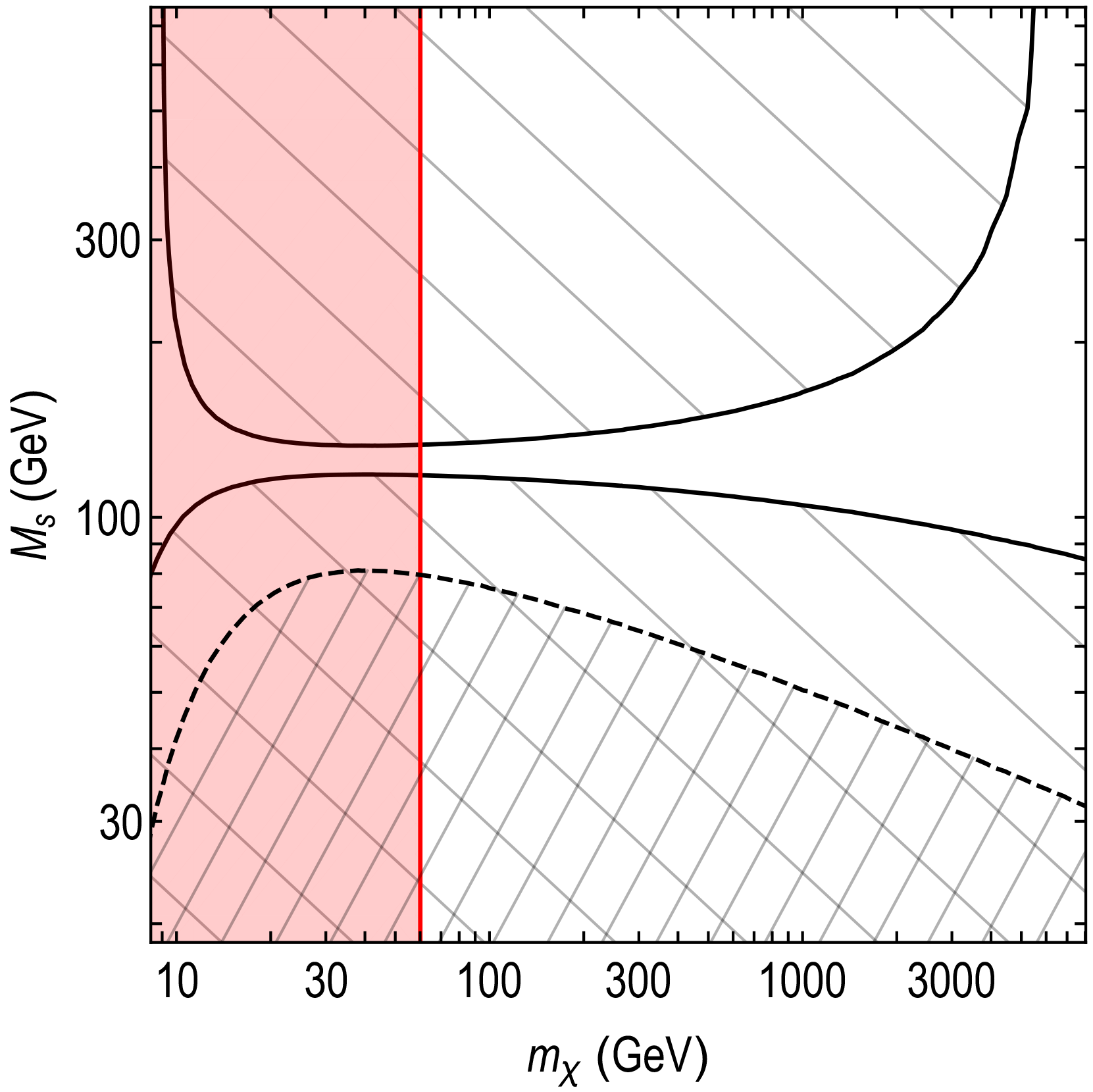}
\caption{Direct detection limits on the H+S model, using LUX
  data. Left panel: limits on the mixing angle $\epsilon$ as a
  function of DM mass for fixed values of the mass of the second
  mediator, for $y_\chi=1$. The solid black line refers to a heavy
  mediator ($M_S=1\TeV$), while the dashed line refers to a mediator
  nearly degenerate with the SM Higgs ($M_S=150\GeV$). The red region
  is excluded by Higgs invisible width constraints, while the blue
  region is excluded by precision electroweak constraints on $\sin
  \epsilon$ from~\citep{Lopez-Val:2014jva}.  Right panel: limits on
  the mass of the second mediator $M_S$ as a function of the dark
  matter mass for fixed values of the mixing angle. The solid black
  line refers to $\sin\epsilon=0.4$, while dashed black line refers to
  $\sin\epsilon=0.03$. The red region is excluded by the Higgs
  invisible width for $\sin\epsilon=0.4$; there is no corresponding
  bound on $m_\chi$ for $\sin\epsilon=0.03$.}
    \label{fig:higgsp}
\end{figure}

DD constraints for these models are generated using tools
from~\citep{DelNobile:2013sia}.  The left panel of
Fig.~\ref{fig:higgsp} shows limits from LUX~\citep{Akerib:2013tjd} on
the mixing parameter $\epsilon$ as a function of the DM mass for
$y_\chi=1$. The solid lines assume a heavy scalar mediator
($M_S=1\TeV$), while the dashed lines refer to a nearly degenerate
scalar of $M_S = 150 \GeV$. The region excluded by invisible Higgs
decays is also shown (red shaded region), which is complementary to
DD, covering the low DM mass region. Finally, the blue shaded region
refers to $\sin\epsilon>0.4$, which is excluded by EW precision
data~\citep{Lopez-Val:2014jva}.  In the right panel we instead fix the
mixing angle to be $\sin\epsilon=0.4$ for the solid lines, and
$\sin\epsilon = 0.03$ for the dashed lines, and show limits on $M_S$
as a function of the DM mass. For $\sin\epsilon=0.4$, one can see that
there is a very large range of DM masses for which only the
nearly-degenerate scenario is allowed.  This permits the second
mediator mass to lie close to that of the 125 GeV Higgs, with larger
and smaller values both excluded. 
This is because the nucleon operator coefficient in eq.~(\ref{eq:nucleonSH}) is very large when mediated by the Higgs boson -- hence to avoid DD constraints, we require the second scalar to have a similar mass in order to have strong cancellation between the two diagrams.  For such a mixing angle, the low DM
mass mass region, $m_\chi \lesssim M_h/2$, is excluded by Higgs
invisible decays (red shaded). For a much smaller scalar mixing angle,
$\sin\epsilon=0.03$, the upper bound on the mass of the second
mediator disappears, opening up the heavier-mediator region.
Note that, as in this plot we are fixing the mixing angle to some fixed value, the perturbativity of the coupling $\lambda_{HS}$ together with eq.~(\ref{eq:mixepsilon}) will necessarily give an upper bound on the mass of the new scalar. Because of this, it's never possible to completely decouple the second scalar by taking the limit $M_S\rightarrow\infty$ while keeping the mixing angle value fixed.

\subsection{Direct detection constraints for 2HDM+S scenarios}

The direct detection formalism for this case is similar to the
preceding one,  however the phenomenology is different because the
mass of the first scalar is not fixed to be the SM Higgs mass and the
mixing angle is unconstrained from SM physics. Moreover, the Yukawa
couplings of the second doublet are no longer forced to all be
rescaled by the same factor, and we will check the effect of the
non-standard Yukawa patterns, as described in the previous sections,
including destructive interference between different quarks types.

The quark and gluon operators are now
\bea
O_q &=& c_q O_1^q = \frac{\lambda_q y_\chi \cos\theta\sin\theta}{\sqrt{2}}\left(\frac{1}{M_{S_1}^2}-\frac{1}{M_{S_2}^2}\right)\bar{\chi}\chi \bar{q}q, \\
O_g &=&  -\frac{c_q}{m_q} O_1^g, 
\eea
where the value of $\lambda_q$ depends on the specific model.  For the
specific Yukawa structures discussed in section~\ref{sec:yukawa}, we
obtain the following coefficients for the nucleon operator:\\
\small
\bea
c_N^{\text{type I}} &=&  m_N\frac{y_\chi \cos\theta\sin\theta}{v\tan\beta}\left(\frac{1}{M_{S_1}^2}-\frac{1}{M_{S_2}^2}\right) \left(\sum_{q=u,d,s} f_{T_q}^N  + \frac{2}{9} f_{T_g} \right),\label{eq:nucleoncoeffTI}\\
c_N^{\text{type II}} &=&  m_N\frac{y_\chi \cos\theta\sin\theta}{v}\left(\frac{1}{M_{S_1}^2}-\frac{1}{M_{S_2}^2}\right) \left(f_{T_u}^N\cot\beta - \tan\beta\sum_{q=d,s} f_{T_q}^N + \frac{2}{9} f_{T_g} \frac{2\cot\beta-\tan\beta}{3} \right), \label{eq:nucleoncoeffTII}\\
c_N^{\text{aligned}} &=&  m_N\frac{y_\chi \cos\theta\sin\theta}{v}\left(\frac{1}{M_{S_1}^2}-\frac{1}{M_{S_2}^2}\right) \left(f_{T_u}^N\tan\gamma_u + \tan\gamma_d\sum_{q=d,s} f_{T_q}^N + \frac{2}{9} f_{T_g} \frac{\tan\gamma_u+2\tan\gamma_d}{3} \right)\label{eq:aligncoeff}, \ \ \ \ \ \\
c_N^{\text{2gen}} &=&  m_N\frac{y_\chi \cos\theta\sin\theta}{v}\left(\frac{1}{M_{S_1}^2}-\frac{1}{M_{S_2}^2}\right) \left(A \left(\frac{f_{T_u}^N}{y_u} + \frac{2f_{T_g}}{27y_c}\right) + B\sum_{q=d,s} \frac{f_{T_q}^N}{y_q} \right).\label{eq:nucleoncoeff2gen}
\eea
\normalsize
One can indeed note the presence of negative interference in the Type
II scenario, and the possibility to also achieve it in the Aligned and
2-generation models, for appropriate values of $\gamma_{u,d}$ and
$\{A,B\}$, respectively.

 \begin{figure}[!ht]
    \centering
    \includegraphics[width=0.45\textwidth]{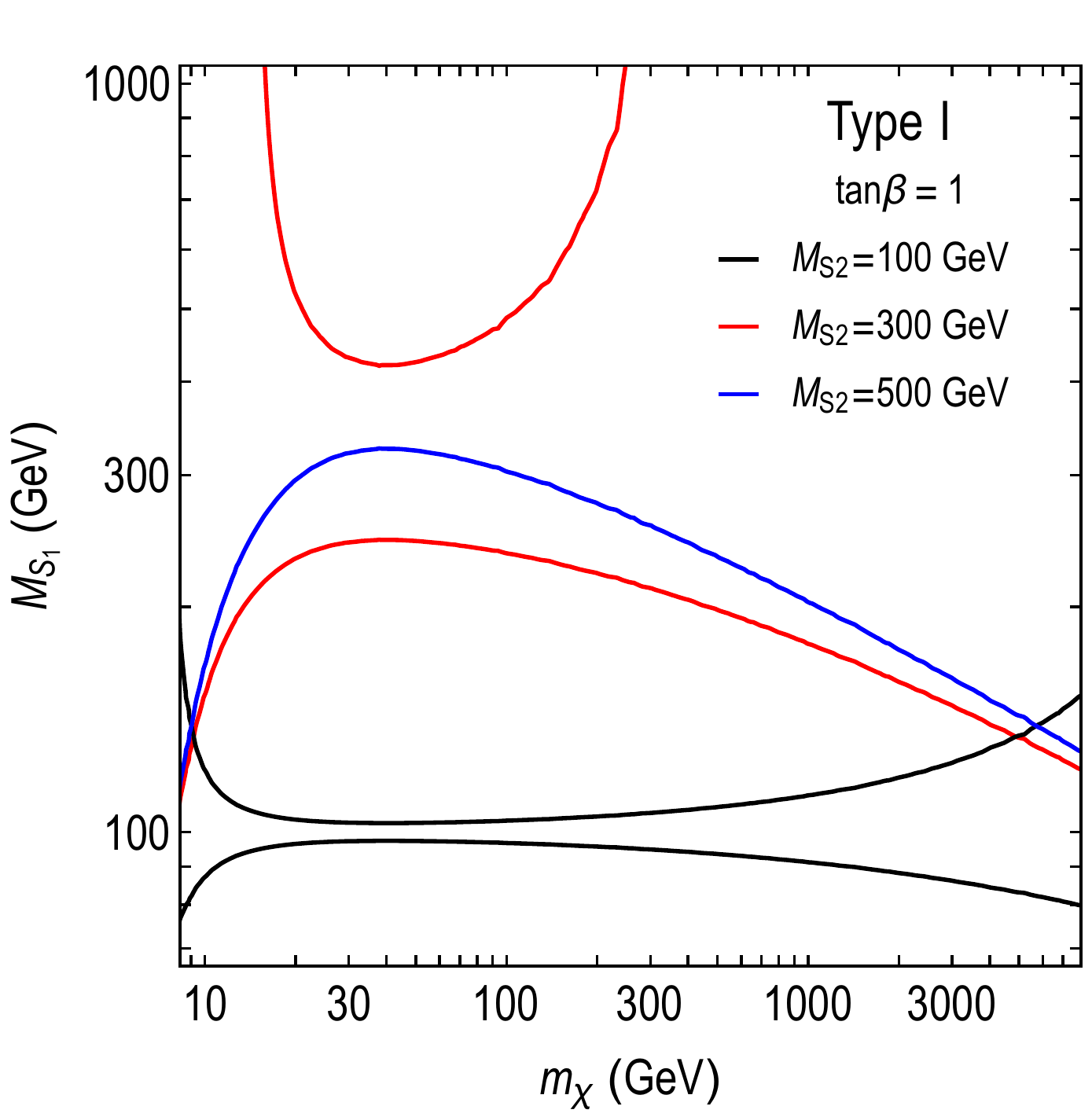} \hspace{1em}
    \includegraphics[width=0.45\textwidth]{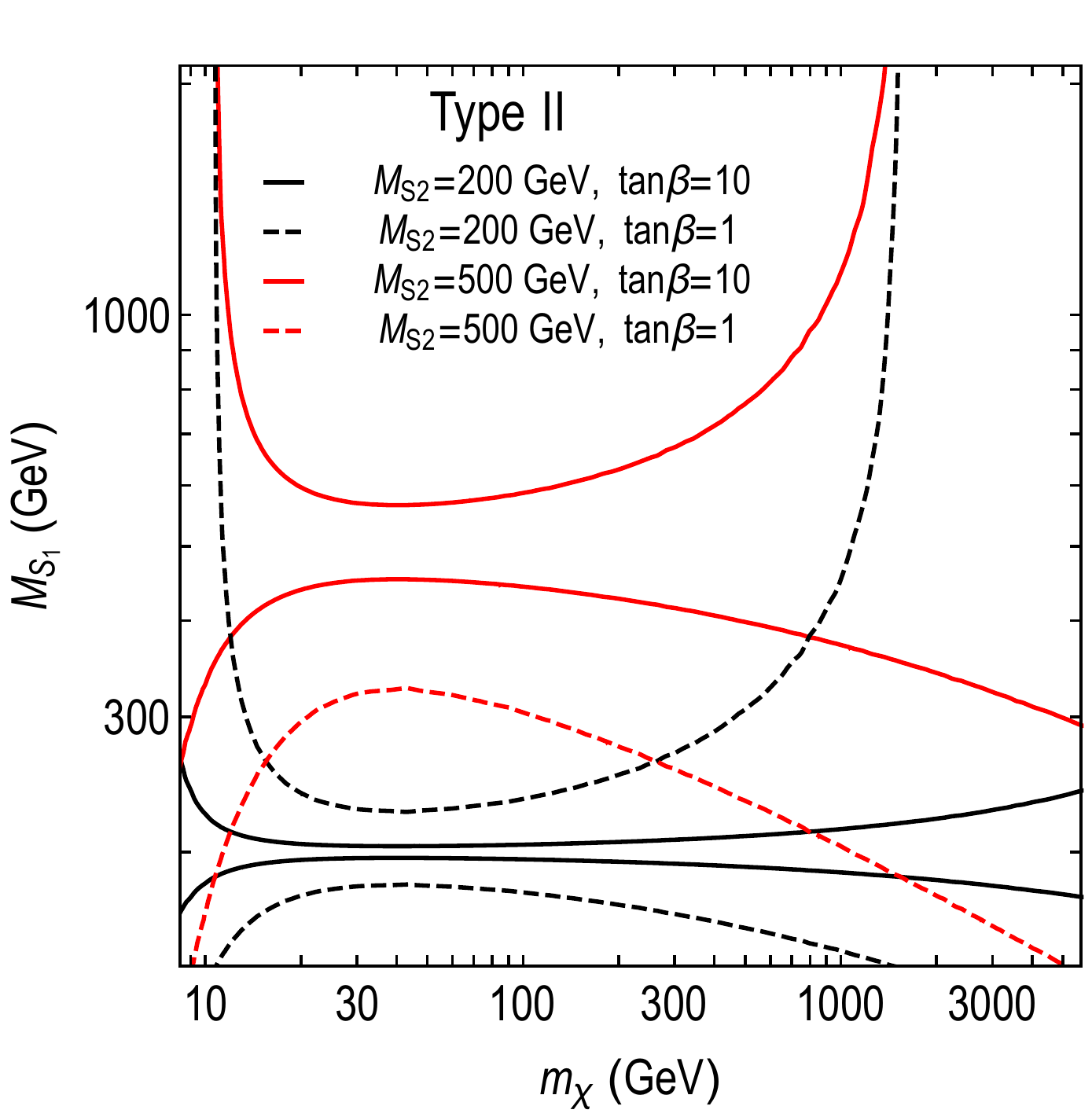}\\
    \includegraphics[width=0.45\textwidth]{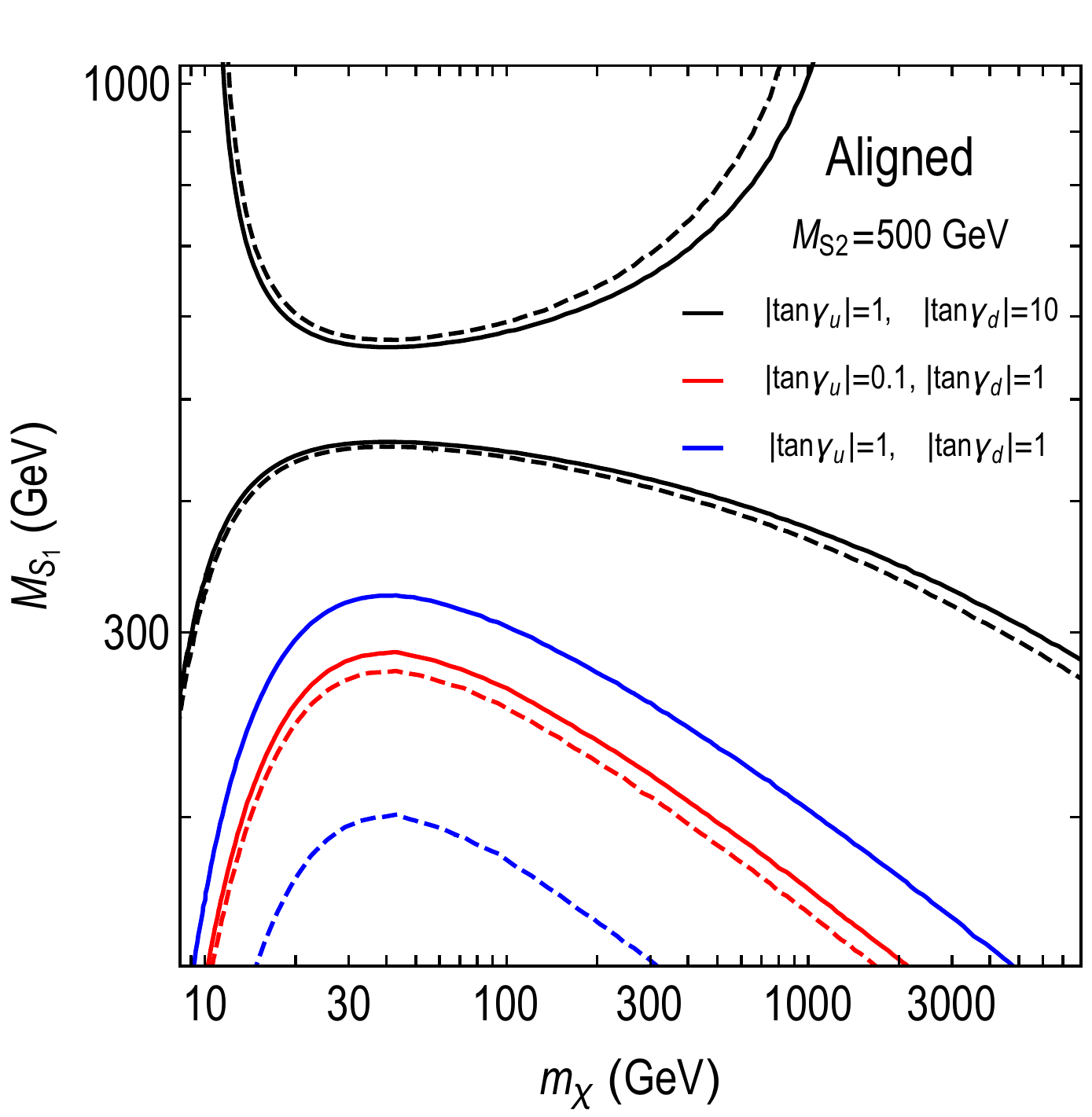} \hspace{1em}
    \includegraphics[width=0.45\textwidth]{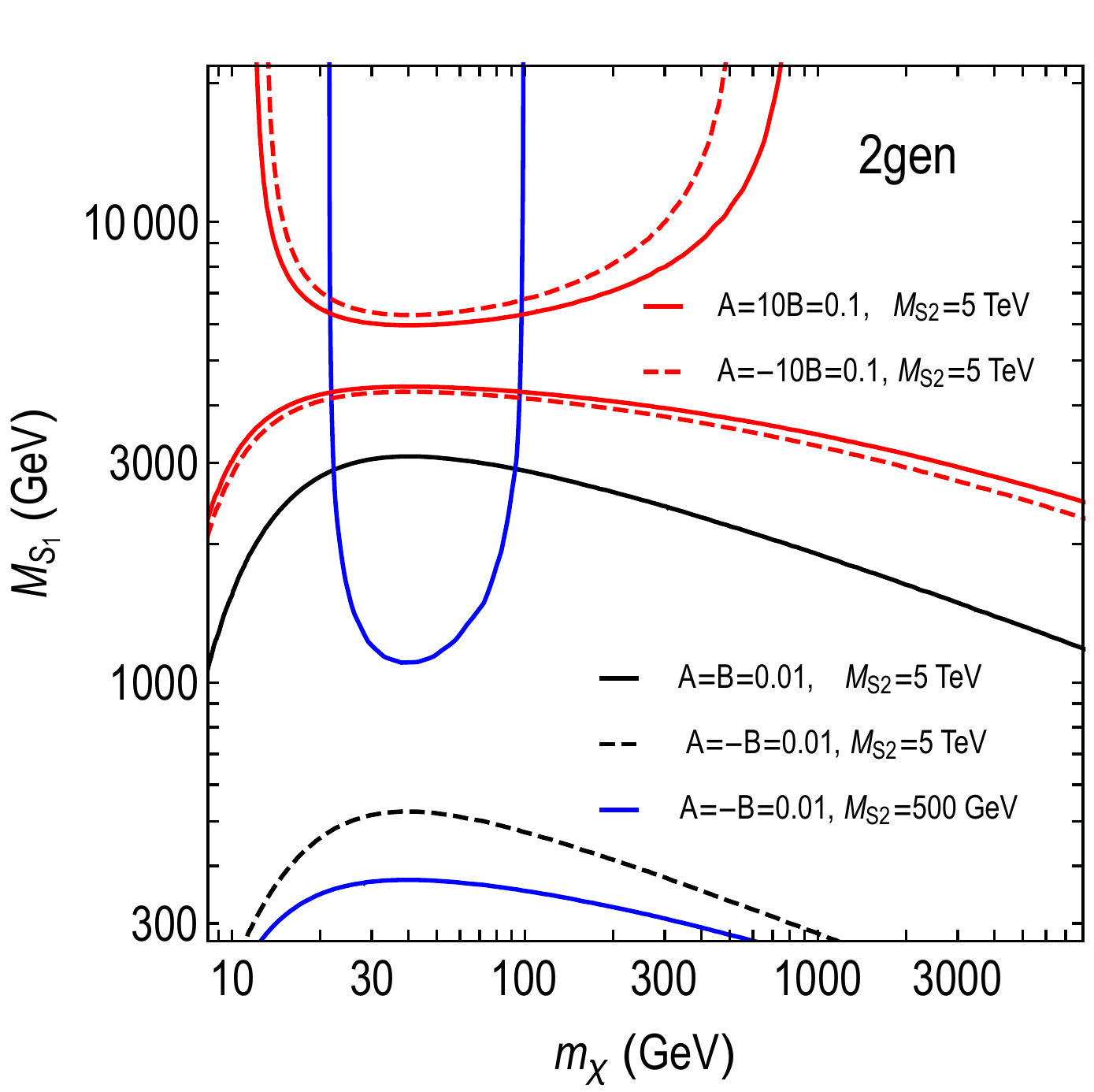}
    \caption{Direct detection limits for Type I, Type II, Aligned and
      2gen 2HDMs, using LUX data. All panels assume maximal mixing
      between the 2 additional scalars ($\theta=\pi/4$). For the
      Aligned model, the solid and dashed curves denote refer to
      $\tan\gamma_u$ and $\tan\gamma_d$ of the same or opposite sign,
      respectively. For the 2-gen model with $A=B=0.01$, $M_{S2}=500$
      GeV or $A=\pm 10B = 0.1$, $M_{S2}=500$ GeV, all the parameter
      space is excluded except for the very degenerate region $M_{S1}
      \approx M_{S2}$, and so was not plotted. }
    \label{fig:2hdmDD}
\end{figure}

Fig.~\ref{fig:2hdmDD} shows the DD limits for $\theta=\pi/4$ for Type
I (upper left panel), Type II (upper right panel), aligned (lower left
panel) and 2-gen (lower right panel) 2HMD+S scenarios. For Type I and
II we choose $\tan\beta=1$ or $1,10$ respectively, while for the
aligned and 2-gen cases we choose different values of $\gamma_{u,d}$
and $A,B$ respectively (including opposite signs). Different value of
$M_{S_2}$ were chosen in each case, as labelled on the figure
caption. A general feature of these results is the presence of not
only a lower bound, but also of an upper bound on $M_{S_1}$, for fixed
values of $M_{S_2}$ below a certain threshold value.

This feature was present also in the H+S model, and its origin is the same; when one of the two mediators has a mass that is too low and would be excluded by DD constraints -- if taken alone -- one can always evade DD constraints by taking the other mediator to be nearly degenerate, thus allowing a cancellation in the nucleon operator coefficients in eq.~(\ref{eq:nucleoncoeffTI}, \ref{eq:nucleoncoeffTII}, \ref{eq:aligncoeff}, \ref{eq:nucleoncoeff2gen}). As before, in these plots we are fixing the mixing angle and the mass of one of the scalars, thus setting an upper bound on the mass of the second scalar due to perturbativity. In this case, however, the upper bound can be increased by having a sufficiently large enough singlet vev $v_s$.

In Type I, if one of the two mediators is light, $M_{S_2} \lesssim
100\GeV$, only the nearly-degenerate scenario is allowed, while
progressively increasing the mass of one of the scalar opens up more
 parameter space in the plane $M_{S_1}-m_\chi$. For
$M_{S_2}=300\GeV$, one still has an upper bound on $M_{S_1}$ in a
certain range of DM masses.  Only when $M_{S_2} \gtrsim 400\GeV$ is
the other mediator allowed to be arbitrarily heavy for any DM mass.
In Type II we note the effect of interference for $\tan\beta=1$
and a fixed value of $M_{S_2}$. While in Type I increasing the value
of $\tan\beta$ leads to weaker limits, for Type II we have the
opposite situation, and this is not only because the coupling to down
quarks increases, but also because we have an important interference
effect for $\tan\beta \lesssim 1$.  For both the mediator masses shown,
$M_{S_2}=200\GeV,$ $500\GeV$ (black, red), the choice $\tan\beta=1$ (dashed) always
results in weaker limits than $\tan\beta=10$ (solid).

An interference effect also arises in the aligned model, where we have
more freedom with the couplings. In this case interference arises when
$\tan\gamma_d \cot \gamma_u<0$. The coefficient in
Eq.~(\ref{eq:aligncoeff}) cancels when $\tan\gamma_d$ and $\tan\gamma_u$
are of similar magnitude but opposite sign.  Of the cases displayed,
we can indeed notice that interference plays an important role only
for $\tan\gamma_u=-\tan\gamma_d=\pm 1$ (blue curves). Finally, in the
model were we couple only the first two generations with equal
couplings, we no longer have contributions from $t,b$ quarks, but
light quarks contributions can be significantly enhanced. Choosing
couplings $A=0.1, B=0.01$ results in more than an order-of-magnitude
increase for the limits on the scalar mediator mass, compared with the
preceding cases. In this model, we can see that interference plays an
important role only if $A\sim -B$.

\section{Conclusions}
\label{sec:conclusions}

We have considered the class of models in which a fermion DM
candidate, uncharged under the SM gauge group, interacts with SM
fermions via the exchange of an $s$-channel scalar mediator.  The
single-mediator Simplified Model version of this scenario is not gauge
invariant.  Instead, this scenario can be realised via the mixing of a
singlet scalar which couples to DM, with an $SU(2)_L$ doublet scalar which
couples to SM fermions.  If this $SU(2)_L$ doublet scalar is none
other than the SM Higgs, this leads to a situation where all
couplings of quarks and leptons to the two mixed scalars are
proportional to the SM Yukawa couplings, with a universal
proportionality constant set by the scalar mixing angle.

Alternatively, the singlet scalar may mix with an additional SU(2)
doublet in a two Higgs doublet model.  This permits much greater
freedom for the masses and mixing of the scalar mediators, and their
Yukawa couplings to SM fermions.  In this case, the proportionality
constant can be made sector dependent, taking different values for the
$u$-quark, $d$-quark and lepton sectors. For example, it is possible to
enhance the coupling to $d$-type quarks, suppress the coupling to
top-quarks, or obtain a leptophilic or leptophobic model. Moreover, it
is possible to obtain flavour dependent couplings without inducing
large FCNCs.  Such Yukawa structures would lead to very different
collider phenomenology.

We calculated direct detection constraints on the H+S and 2HDM+S
models, which both feature spin-independent DM-nucleon scattering.
Much of the parameter space of the H+S model is excluded by direct
detection data, unless either the scalar mixing angle is very small,
or the mass of the new scalar is approximately degenerate with the SM
Higgs, suppressing the scattering cross section via destructive
interference.  For the 2HDM+S model, interference of the two scalar
mediators is again very important.  Moreover, an additional source of
interference is now provided by relative cancellations between
different quarks in the nucleus.  This occurs when up and down type
quarks have Yukawa couplings of opposite sign, as can occur in the
Type II, Aligned and 2-generation Yukawa structures we examined. As a
result, unlike the H+S model, the 2HDM+S scenario has substantial
parameter space that is not eliminated by direct detection.

It is clear that the phenomenology of these minimal self consistent
scenarios (both H+S and 2HDM+S) is much richer than that of the
single-mediator (non gauge invariant) Simplified Model.  If there is a
hierarchy of scalar masses, such that the heaviest decouples and
interactions are dominated by the exchange of the lightest, we recover
the structure of the single-mediator model from a gauge invariant
framework.  In general, all three scalars of the 2HDM+S scenario would
mix.  However, we took a generalised Higgs alignment limit, where the
SM Higgs decouples and the singlet mixes only with the additional
doublet. Experimental constraints on Higgs properties require that
this limit is at least approximately realised, though there would be
scope to relax this assumption while still abiding by current Higgs
constraints.  Regardless, it is evident that the $s$-channel scalar
mediator scenario should in general be analysed in a multi-mediator
context, with a minimum of two scalars.

\section*{Acknowledgements}
This work was supported in part by the Australian Research Council.
We thank Yi Cai, Matthew Dolan, Francesco D'Eramo and Gino Isidori
for helpful discussions, and Ulrich Haisch for detailed comments on an
earlier draft of the manuscript. Feynman diagrams were drawn using
{\sc TikZ-Feynman} \cite{Ellis:2016jkw}.

\appendix
\section{Unitarity, Stability and perturbativity of 2HDM+S}
\label{sec:s2hdmconditions}

\subsection{Unitarity and perturbativity}
\label{sec:unitarity}

\begin{figure}[h]\centering
	\begin{tikzpicture}[scale=1.2]
	\draw[->] (-2,0) -- (4,0);
	\draw[->] (0,-0.5) -- (0,2.5);
	\draw (1.5,3pt) -- (1.5,-3pt)   node [below]  {$\tfrac{T}{2}$};
	\draw (3,3pt) -- (3,-3pt)   node [below] {$T$};
	\draw (3pt,2) -- (-3pt,2)   node [left]   {$i$};
    \draw (3pt,1) -- (-3pt,1)   node [left] {$\tfrac{i}{2}$};
    \path [draw=black,fill=gray,semitransparent] (0,1) circle (1);
     \draw [dashed,red,thick,domain=0:180] plot ({1.5+1.5*cos(\x)}, {1.5*sin(\x)});
     \draw [dotted,red,thick,domain=0:180] plot ({2*cos(\x)}, {2*sin(\x)});
     \draw (2,3pt) -- (2,-3pt)   node [below] {$1$};
     \draw (-2,3pt) -- (-2,-3pt)   node [below] {$-1$};
     \draw (1/8/3.14,3pt) -- (1/8/3.14,-3pt)   node [below] {$\tfrac{1}{16\pi}$};
    \draw [dashed,thick] (3,0) -- (0,2);
    \draw [dotted] (0,0) -- ({12/13,18/13}) node [above,yshift=1mm,xshift=2mm] {$T_U$};
    \path [fill=red]  ({12/13,18/13}) circle (0.07);
    \path [fill=black]  (3,0) circle (0.04);
    \path [fill=black]  (0,2) circle (0.04);
    \path [fill=black]  (0,0) circle (0.04);
\end{tikzpicture}
\caption{Argand circle and Thales projection $T_U$.}
\label{fig:argand}
\end{figure}
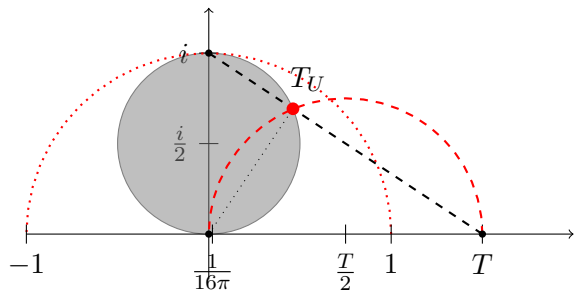%

We start with the scattering amplitude $M$ for the scalar particles,
\bea
M_{\Phi_i\Phi_j\rightarrow\Phi_k\Phi_l} = \frac{\partial^4 V}{\partial \Phi_i \partial \Phi_j \partial \Phi_k^\dagger \partial \Phi_l^\dagger}. 
\eea
This matrix is hermitian, so its eigenvalues are real. The $T$-matrix is related to it by 
\bea
T = \frac{1}{16\pi} M.
\eea
The unitarity of the model is satisfied as long as the eigenvalues of $T$ satisfy\footnote{See \citep{Bell:2016obu} for more details.} 
\bea
|\text{Eigenvalues}[T]| \le 1.
\eea
This condition means that the scattering amplitude at tree level does not exceed the unitarity limit for the cross section. Looking at the Argand circumference in Fig.~\ref{fig:argand}, this means that the amplitude $T$ lies inside the red dotted circumference of radius $1$.
If, however, we also want perturbativity, then we need to impose a more stringent condition \citep{Kanemura:2015ska}
\bea
|\text{Eigenvalues}[M]| \le \xi \rightarrow |\text{Eigenvalues}[T]| \le \frac{\xi}{16\pi},
\eea
where the choice of a maximum value of $\xi=1,2$ is somehow arbitrary, and this condition means that our amplitude, that is real and so is lying on the real axes, is not far away from the unitarity circle, and then the loop corrections necessary to move it inside it are small compared to the tree level amplitude\footnote{Or that, in other words, the Thales projection $T_U$ is very close to the tree level amplitude $T$, namely $|T_U-T|^2\ll |T|^2$}.

\begin{figure}[!t]
    \centering
    \includegraphics[width=0.45\textwidth]{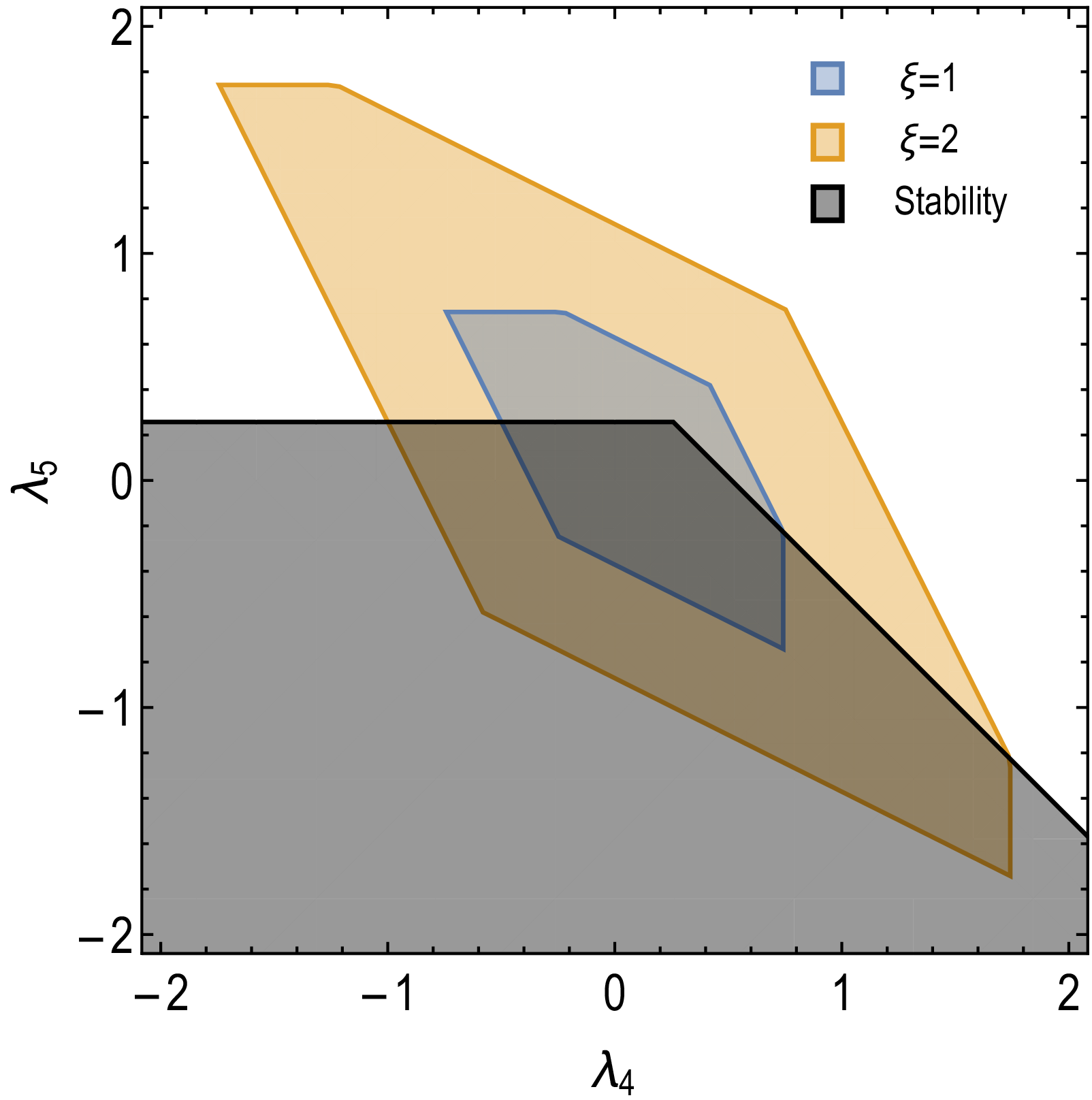}\hspace{1em}
    \includegraphics[width=0.47\textwidth]{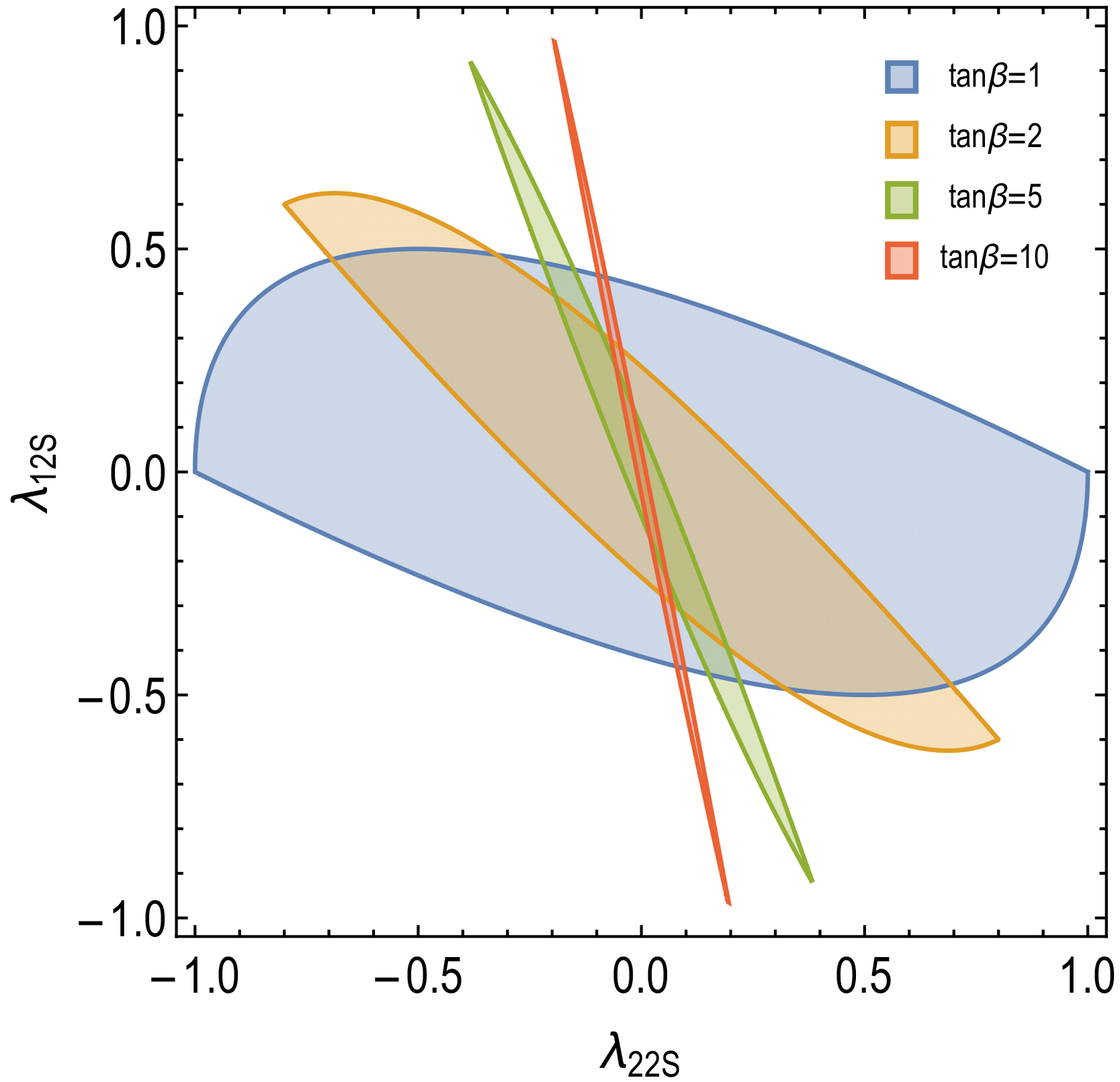}
        \caption{Regions of the parameters space satisfying perturbativity constrains from scattering matrix $A_{0+\rightarrow0+}$. Left panel: the coloured regions are where the model is perturbative, while the black shaded region is the one excluded by global stability of the potential (CP2 symmetry assumed in this plot). Right panel: Allowed regions for $\xi = 1$ for varying $\tan \beta$. Note that the dimensions of all regions in this plot scale linearly with $\xi$.}
    \label{fig:perturb1}
\end{figure}

The scattering matrix $M$ is block diagonal, the blocks being related to initial and final states of same electric charge. 
The charge $1$ block can be put in block diagonal form as
\bea
A_{0+\rightarrow0+} &=& \text{Block}[B_{1},B_{2},B_{3}, D(\lambda_3+\lambda_5,\lambda_3-\lambda_5,\lambda_3+\lambda_4,\lambda_3-\lambda_4)],\\
B_{1} &=& \left(
\begin{array}{cc}
 \lambda_{11S} & \lambda_{12S} \\
 \lambda_{12S} & \lambda_{22S} \\
\end{array}
\right),\\
B_{2} &=& \left(
\begin{array}{cc}
 \lambda _{1} & \lambda _{4} \\
 \lambda _{4} & \lambda _{2} \\
\end{array}
\right),\\
B_{3} &=& \left(
\begin{array}{cc}
 \lambda _{1} & \lambda _{5} \\
 \lambda _{5} & \lambda _{2} \\
\end{array}
\right).
\eea
The charge $2$ scattering matrix is instead
\bea
A_{++\rightarrow ++} &=& \text{Block}[B_{2},D(\lambda_3+\lambda_4)]\label{eq:scattpp}.
\eea
From $A_{0+ \to 0+}$ and $A_{++ \to ++}$ we get the following constraints:
\bea
|\lambda _3|+|\lambda _4|< 1,\\
|\lambda _3|+|\lambda _5|< 1,\\
\lambda _1+\lambda _2+\sqrt{\lambda _1^2-2 \lambda _2 \lambda _1+\lambda _2^2+4 \lambda
   _5^2}< 2,\label{eq:pert1}\\
   \lambda _1+\lambda _2+\sqrt{\lambda _1^2-2 \lambda _2 \lambda _1+\lambda _2^2+4 \lambda
   _4^2}< 2,\label{eq:pert2} \\
   \lambda_{11S} + \lambda_{22S} + \sqrt{\lambda_{11S}^2 + 4 \lambda_{12S}^2 - 2 \lambda_{11S} \lambda_{22S} + \lambda_{22S}^2} < 2,
\eea
where \eqref{eq:pert1} and \eqref{eq:pert2} can be rewritten as
\bea
\lambda_{4,5}^2 < (1-\lambda_1)(1-\lambda_2), \quad \lambda_{1,2}<1.
\eea
Regions of the parameter space where eigenvalues of $B_2,B_3$ are perturbative are shown in the left panel of Fig. \ref{fig:perturb1} for $\xi=1,2$, assuming the presence of the CP2 symmetry\footnote{Note that this condition becomes $\lambda_1 = \lambda_2 = \frac{1}{2} (\lambda_3 + \lambda_4 + \lambda_5)$ when the normalization of the Lagrangian is taken to be the alternative of what we have chosen, as is done in \cite{Dev:2015bta}.} ($\lambda_1=\lambda_2 = \lambda_3 + \lambda_4 + \lambda_5 =\frac{M_h^2}{v^2}$), so that no condition depends on $\beta$. The right panel instead show the regions of parameters space where eigenvalues of $B_{1}$ are perturbative, for several values of $\tan\beta$ and $\xi=1$ (note that the regions in this plot scale linearly with $\xi$). At large $\tan\beta$ the parameter space shrinks due to the alignment condition, and fine-tuning is necessary to guarantee the unitarity and perturbativity of all couplings.

 \begin{figure}[!t]
    \centering
    \includegraphics[width=0.45\textwidth]{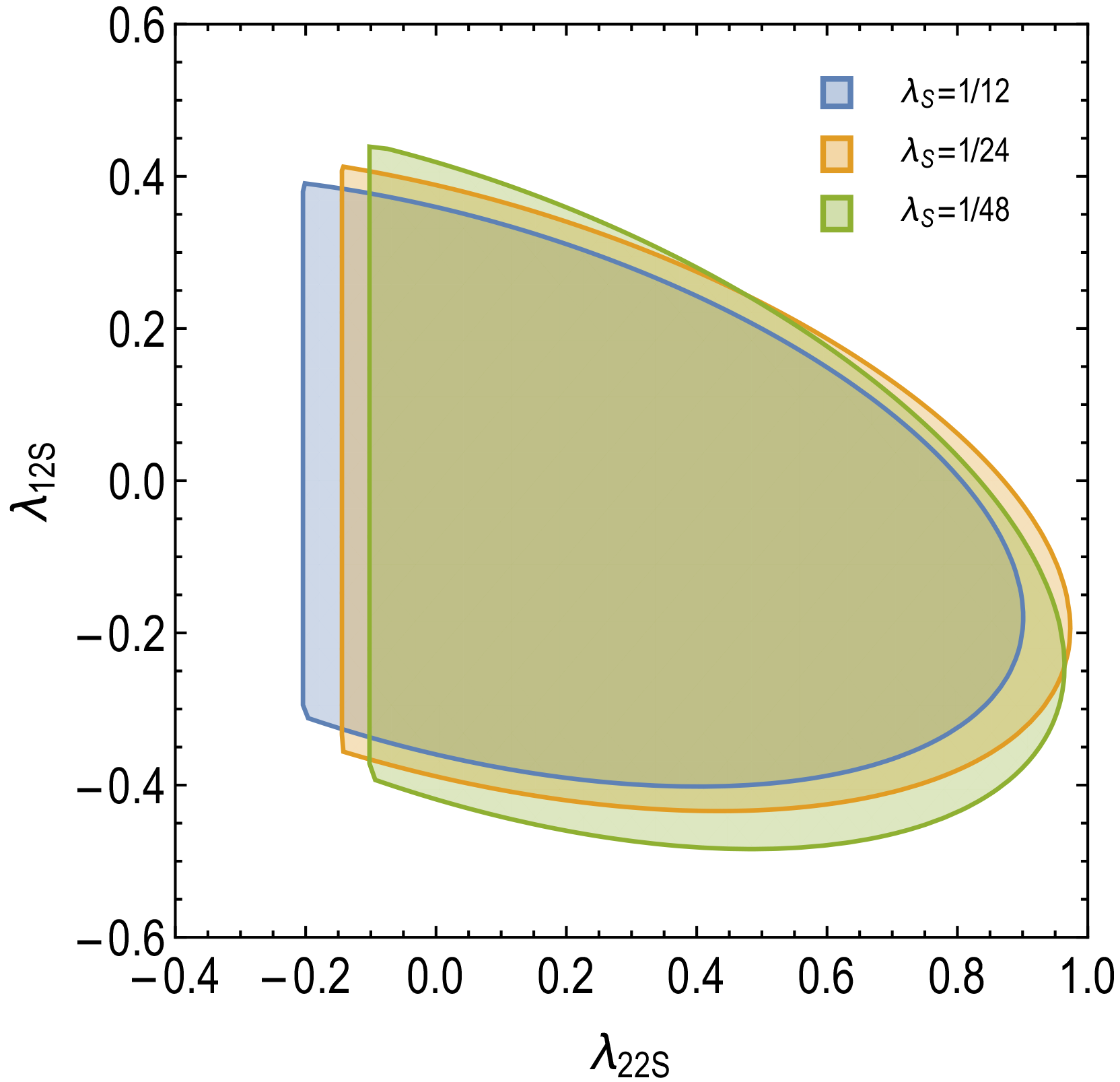}
    \includegraphics[width=0.45\textwidth]{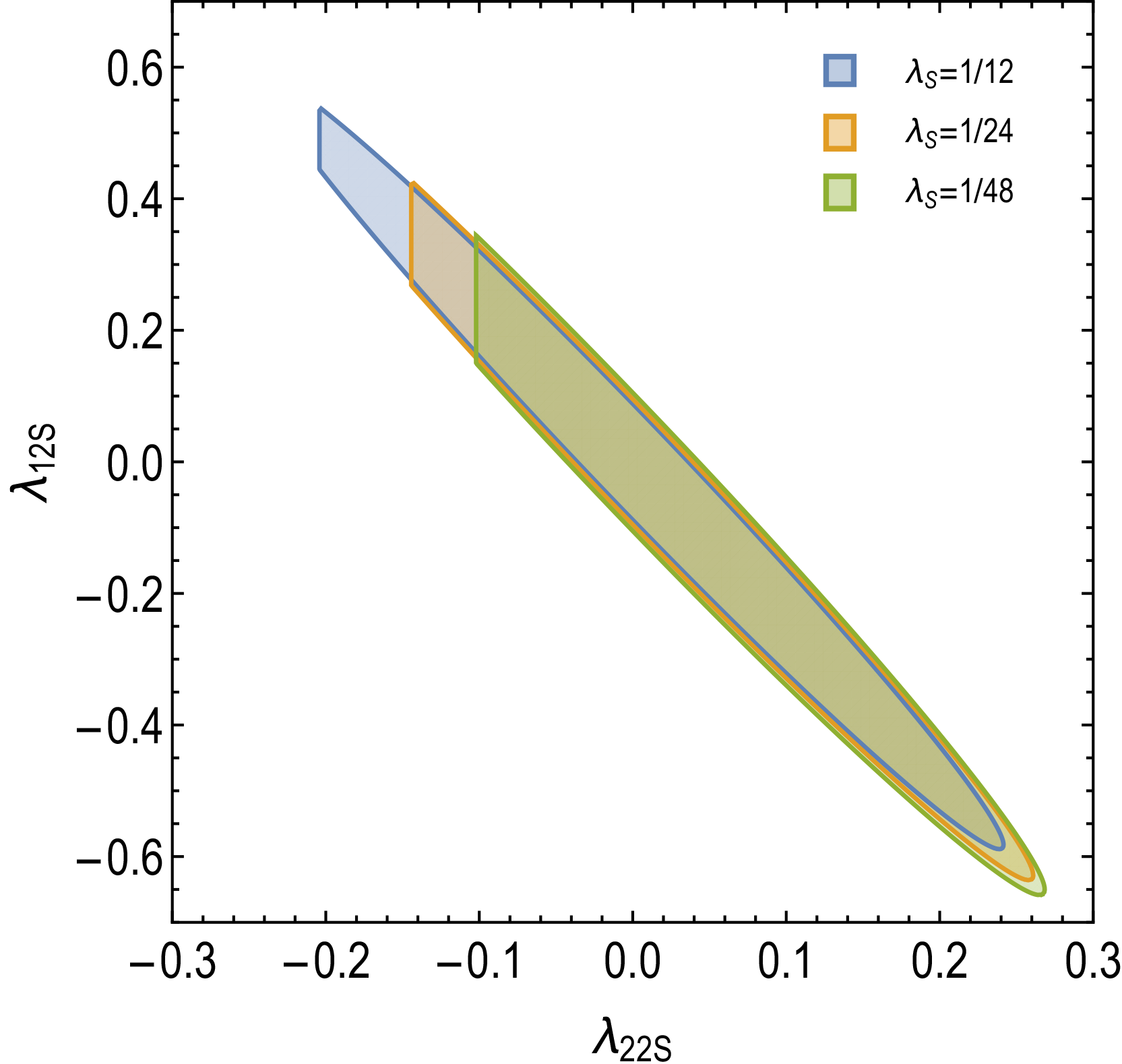}
    \caption{Regions of the parameters space satisfying perturbativity constrains from scattering matrix $A_{00\rightarrow00}$ for $\xi = 2$. Left panel: $\tan\beta=1$, Right panel: $\tan\beta=5$. Lower limits on $\lambda_{22S}$ come from global stability conditions. 
    }
    \label{fig:perturb2}
\end{figure}

The zero-charge scattering matrix is
\bea
A_{00\rightarrow00} &=& \text{Block}[B_1,B_1,B_2,B_3,B_3,B_4, D(\lambda_3+\lambda_5,\lambda_3-\lambda_5,\lambda_3+\lambda_4,\lambda_3-\lambda_4)],\\
B_{4} &=& \left(
\begin{array}{cccc}
 \lambda_{3}+2\lambda_4+3\lambda_5 & 0 & 0 & 2\lambda_{12S} \\
 0 & 3\lambda_1 & 2\lambda_3+\lambda_4 &\sqrt{2}\lambda_{11S}\\
 0 & 2\lambda_3+\lambda_4 & 3\lambda_2 &\sqrt{2}\lambda_{22S}\\
 2\lambda_{12S} & \sqrt{2}\lambda_{11S} &\sqrt{2}\lambda_{22S} & 6\lambda_S\\
\end{array}
\right).
\eea
All eigenvalues of $A_{00 \to 00}$ except the ones of $B_4$ already satisfy perturbativity constraints in the regions where the matrices $A_{++\rightarrow ++},A_{0+\rightarrow 0+}$ do. The eigenvalues of $B_4$ can be checked numerically. Allowed regions in parameter space are show in Fig. \ref{fig:perturb2} for $\tan\beta=1$ (left) and $\tan\beta=5$ (right), for several values of $\lambda_S$. For this plot we only show the regions for $\xi=2$, as very small region of the parameter space satisfies the perturbativity conditions for $\xi=1$\footnote{Varying the value of $\xi$ results in an approximate shrinking of the regions by a $\xi$-dependent factor. The critical point below which the regions shrink to the origin is $\xi=0.9$.}. This happens because of our choice of imposing the CP2 symmetry, removing the freedom of $\lambda_3$. We also set the value of $\lambda_4,\lambda_5$ to the ones that maximise the allowed parameter space.

\begin{figure}[!t]
    \centering
    \includegraphics[width=0.65\textwidth]{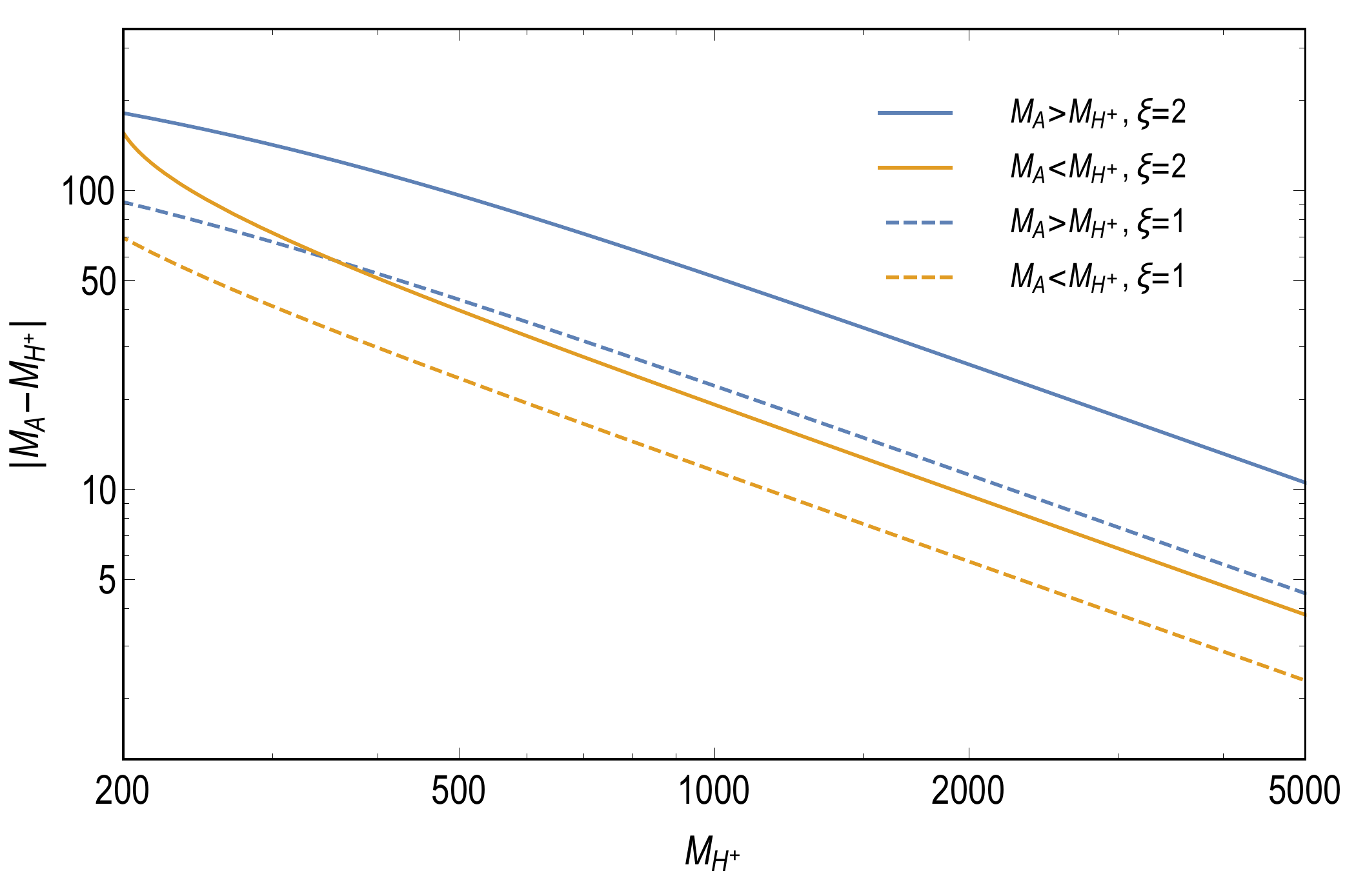}
    \caption{Maximum mass splitting between the charged scalar and the pseudoscalar for different values of $\xi$.}
    \label{fig:masssplitting}
\end{figure}

The perturbativity constraints on $\lambda_{4,5}$ imply that the mass splitting between the charged scalar and the pseudoscalar can have a maximum allowed value, that is shown in Fig. \ref{fig:masssplitting} as a function of the charged scalar mass for $\xi=1,2$.

Note that the mass splitting in the additional doublet affects also EW oblique parameters at loop level. The most stringent constraint comes from the $T$ parameter \citep{Lopez-Val:2013yba} and usually imposes much more stringent constraints on the maximum splittings $\Delta_{A+}=M_A^2-M_{H^+}^2$, $\Delta_{S_{1,2}+}=M_{S_{1,2}}^2-M_{H^+}^2$ and the mixing angle $\theta$ than the ones obtained by the perturbativity limits on the couplings.

All the plots in this section account for the following necessary global stability conditions:
\bea
\lambda_{1,2,S} &>& 0,\\
\sqrt{\lambda_1\lambda_2} &>& -\lambda_{3},\\
\sqrt{2\lambda_{1}\lambda_S} &>& -\lambda_{11S},\\
\sqrt{2\lambda_{2}\lambda_S} &>& -\lambda_{22S},\\
\sqrt{\lambda_1\lambda_2} &>& |\lambda_5| - \lambda_3-\lambda_4.
\eea

\section{RGEs for 2-gen 2HDM+S}
\label{sec:rge2gen}
2HDM bring in the picture a lot of constraints from flavour physics. Even with our MFV approach, there are two main sources of new flavour physics: First, the presence of a new charged scalar \citep{Jung:2010ik}, and second the RGE running of the Yukawa couplings \citep{Cvetic:1998uw}. In models with NFC coming from a $Z_2$ symmetry, the structure of the couplings is stable under quantum corrections and this point does not apply \citep{Ferreira:2010xe}.
Updated constrains for the $Z_2$ and aligned model can be found in \citep{Enomoto:2015wbn}.

The 2-gen model suppresses constraints coming from loop diagrams involving $t,b$ quarks running in the loop and $b,t$ quarks as external particles thanks to it coupling structure. 
We analyse here RGE runnings to check that they are indeed stable.
We can write the RGE equations for a 2HDM, in their most general form, as
\begin{align}
16 \pi^2 \frac{d}{d \ln \mu} {\tilde Y}_k^u(\mu)  
= & {\Bigg \{} 
N_{\mathrm{c}} \sum_{\ell=h,H}
{\mathrm{Tr}} \left[ {\tilde Y}_k^u 
{\tilde Y}_{\ell}^{u\dagger} +
 {\tilde Y}_{\ell}^d {\tilde Y}_{k}^{d\dagger} \right]
{\tilde Y}_{\ell}^u
+\frac{1}{2} \sum_{\ell=h,H} \left[
{\tilde Y}_{\ell}^u {\tilde Y}_{\ell}^{u\dagger} +
{\tilde Y}_{\ell}^d {\tilde Y}_{\ell}^{d\dagger} 
\right] {\tilde Y}_{k}^u  \nonumber\\
& + {\tilde Y}_{k}^u \sum_{\ell=h,H}
{\tilde Y}_{\ell}^{u\dagger} {\tilde Y}_{\ell}^u
-2 \sum_{\ell=h,H} \left[
{\tilde Y}_{\ell}^d {\tilde Y}_{k}^{d\dagger}{\tilde Y}_{\ell}^u
\right]  - A_U {\tilde Y}_{k}^u
{\Bigg \}} \ ,\\
16 \pi^2 \frac{d}{d \ln \mu} {\tilde Y}_k^d(\mu)  
= & 
{\Bigg \{} 
N_{\mathrm{c}} \sum_{\ell=h,H}
{\mathrm{Tr}} \left[ {\tilde Y}_k^d 
{\tilde Y}_{\ell}^{d\dagger} +
 {\tilde Y}_{\ell}^u {\tilde Y}_{k}^{u\dagger} \right]
{\tilde Y}_{\ell}^d
+\frac{1}{2} \sum_{\ell=h,H} \left[
{\tilde Y}_{\ell}^d {\tilde Y}_{\ell}^{d\dagger} +
{\tilde Y}_{\ell}^u {\tilde Y}_{\ell}^{u\dagger} 
\right] {\tilde Y}_{k}^d \nonumber\\
&+ {\tilde Y}_{k}^d \sum_{\ell=h,H}
{\tilde Y}_{\ell}^{d\dagger} {\tilde Y}_{\ell}^d
-2 \sum_{\ell=h,H} \left[
{\tilde Y}_{\ell}^u {\tilde Y}_{k}^{u\dagger}{\tilde Y}_{\ell}^d
\right]  - A_D {\tilde Y}_{k}^d
{\Bigg \}} \ ,
\end{align}
where $k=h,H$ and
\bea
A_U &=& 8 g_3^2+ \frac{9}{4} g_2^2 + \frac{17}{12} g_1^2,\\
A_D &=& A_U - g_1^2.
\eea
We can now rewrite them under the assumption that the Yukawa matrices are approximately of the form
\bea
Y_H^U &=& V^\dagger \left(A P_{12} + a \lambda_{FC}^U\right),\\
Y_H^D &=& B P_{12} + b \lambda_{FC}^D,\\
Y_h^U &=& V^\dagger \left(y_t P_3 + c \lambda_{FC}^U\right),\\
Y_h^D &=& y_b P_3 + d \lambda_{FC}^D,
\eea
where
\bea
\lambda_{FC}^U &=& V P_{12}V^\dagger P_3,\\
\lambda_{FC}^D &=& P_{12}V P_3 V^\dagger ,
\eea
and $a,b,c,d$ are zero at some scale and always very small at any relevant scale. Under this assumption, neglecting $a,b,c,d$ terms on the RHS of the RGE equations, together with terms of order $O \left( y_c,y_s,y_d,y_u, \frac{y_b^2}{y_t^2} \right)$, the set of equations can be solved in sequence:
\bea
16\pi^2 \frac{d y_t}{d\log E} &=& \left(\frac{9}{2}y_t^2-A_U\right)y_t,\\
16\pi^2 \frac{d y_b}{d\log E} &=& \left(\frac{3}{2}y_t^2-A_D\right)y_b,\\
16\pi^2 \frac{d A}{d\log E} &=& \left(\frac{15}{2}A^2+\frac{9}{2}B^2-A_U\right)A,\\
16\pi^2 \frac{d B}{d\log E} &=& \left(\frac{15}{2}B^2+\frac{9}{2}A^2-A_D\right)B,\\
16\pi^2 \frac{d a}{d\log E} &=& \left(\frac{1}{2}y_b^2+\frac{3}{2}B^2\right)A,\\
16\pi^2 \frac{d b}{d\log E} &=& \left(\frac{1}{2}y_t^2+\frac{3}{2}A^2\right)B,\\
16\pi^2 \frac{d c}{d\log E} &=& \left(\frac{1}{2}B^2+\frac{3}{2}y_b^2\right)y_t,\\
16\pi^2 \frac{d d}{d\log E} &=& \left(\frac{1}{2}A^2 +\frac{3}{2}y_t^2\right)y_b.\\
\eea
The running of gauge coupling is described by
\bea
16\pi^2 \frac{d g_i}{d\log E} &=& -C_i g_i^3,\\
C_3 = \frac{1}{3}\left(11N_c-2n_q\right), \qquad C_2 &=& 7-\frac{2}{3} n_q, \qquad C_1=-\frac{1}{3}-\frac{10}{9}n_q,
\eea
and we fix
\bea
\alpha_3(M_Z) &=& 0.118,\\
\alpha_2(M_Z) &=& 0.0332,\\
\alpha_1(M_Z) &=& 0.0101,\\
y_t(\tilde{m}_t) &=& \frac{\sqrt{2}\tilde{m}_t}{v}, \qquad \tilde{m}_t = 166 \GeV,\\
y_b(m_b) &=& \frac{\sqrt{2}m_b}{v}, \qquad m_b = 4.3 \GeV.
\eea
Note that $\tilde{m}_t$ has been chosen as in \citep{Cvetic:1998uw} and
\bea
A(\tilde{m}_t) &=& 0.1,\\
B(\tilde{m}_t) &=& 0.01,\\
a(\tilde{m}_t) &=& b(\tilde{m}_t) = c(\tilde{m}_t) = d(\tilde{m}_t) = 0.
\eea
 \begin{figure}[!t]
    \centering
    \includegraphics[width=0.65\textwidth]{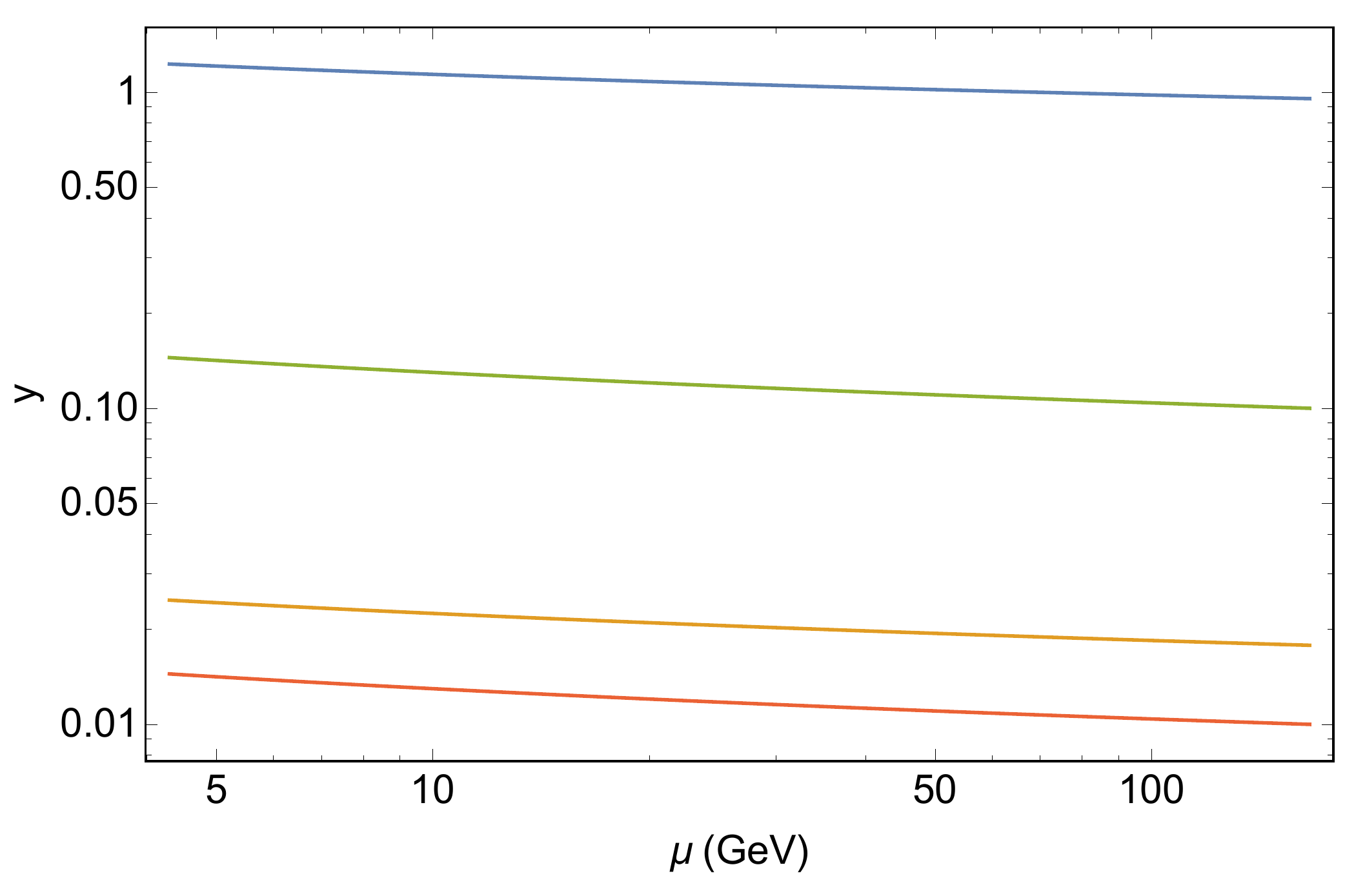}
    \caption{Yukawa couplings as a function of the energy scale $\mu$. From top to bottom: $y_t, A, y_b, B$. With these initial values the hypothesis $A^2\ll y_t^2$ and $B^2 \ll y_b^2$ is valid at all energy scales.}
    \label{fig:rge2gen}
\end{figure}
The behaviour of the couplings is shown in Fig. \ref{fig:rge2gen}. The resulting values of the flavour-violating coefficients are approximately\footnote{Note that if $A$ and $B$ are sufficiently small, equations for $A,B,a,b$ are linear in $A,B$  and equations for $c,d$ are linear in $A^2,B^2$, thus for any starting value such that $A^2\ll y_t^2, B^2 \ll y_b^2$, this is a suitable approximation.}
\bea
|a(m_b)| &=& 1.2 \times 10^{-5} A(\tilde{m}_t),\\
|b(m_b)| &=& 1.6 \times 10^{-2} B(\tilde{m}_t),\\
|c(m_b)| &=& 1.8 \times 10^{-2} B^2(\tilde{m}_t) + 5.4 \times 10^{-6},\\
|d(m_b)| &=& 3.5 \times 10^{-4} A^2(\tilde{m}_t) + 7.4 \times 10^{-4}.
\eea
Of these, the most dangerous contribution is the coefficient $b(m_b)$, that generates $b\rightarrow s$ FCNC. Using the limit taken from \citep{Crivellin:2013wna}, one has 
\bea
\frac{b(m_b)}{0.01}\times 500\GeV\sqrt{\frac{\cos^2\theta}{M_{S_1}^2}+\frac{\sin^2\theta}{M_{S_2}^2}}\lesssim 3,
\eea
which is always satisfied under our assumption $B(\tilde{m}_t)\lesssim y_b$.

\section{Near-alignment and the $V_{\text{\sc ckm}}$ pattern}
\label{sec:ckmexplanation}
We now decide to work in a different framework. We can drop the alignment condition we imposed in Sec. \ref{sec:2hdmmassspectrum}, and instead suppose only an approximate alignment. This means, in the usual 2HDM formalism, that
\bea
\left(
\begin{array}{c}
\rho_1   \\
\rho_2  \\
\end{array}
\right) &=& R\left(\frac{\pi}{2}-\alpha\right) \left(
\begin{array}{c}
h   \\
H  \\
\end{array}
\right),
\eea
where $h,H$ are the two mass eigenstates, and $h$ is the 125 GeV SM Higgs boson. In this case the SM Higgs $h_{\text{\sc sm}}$ will not be a mass eigenstate, like when we have perfect alignment, but rather a mixture, as it is related to the Higgs basis by a rotation
\bea
\left(
\begin{array}{c}
h_{\text{\sc sm}}  \\
H_\perp  \\
\end{array}
\right) &=& R(\beta) \left(
\begin{array}{c}
\rho_1  \\
\rho_2  \\
\end{array}
\right)= R(\beta)R\left(\frac{\pi}{2}-\alpha\right) \left(
\begin{array}{c}
h \\
H  \\
\end{array}
\right)= R\left(\beta-\alpha+\frac{\pi}{2}\right) \left(
\begin{array}{c}
h \\
H  \\
\end{array}
\right),
\eea
where $H_\perp$ is the linear combination orthogonal to the SM one.
We can then rewrite Eq.~(\ref{eq:alignh}), (\ref{eq:alignH}) as
\bea
\Phi_h = \cos\beta \Phi_1 + \sin\beta \Phi_2 = \left(
\begin{array}{cc}
 G^+ \\
\frac{1}{\sqrt{2}} \left(v + \sin(\alpha-\beta)h_{\text{\sc sm}} -\cos(\alpha-\beta)H_\perp + G^0 \right) \\
\end{array}
\right),\label{eq:nalignh}\\
\Phi_H = -\sin\beta \Phi_1 + \cos\beta \Phi_2 = \left(
\begin{array}{cc}
 H^+ \\
\frac{1}{\sqrt{2}} \left( \cos(\alpha-\beta)h_{\text{\sc sm}} + \sin(\alpha-\beta)H_\perp + A \right) \\
\end{array}
\right).\label{eq:nalignH}
\eea
The near-alignment will be defined as
\bea
0 \neq \cos(\beta-\alpha)\ll 1 \rightarrow \alpha-\beta-\pi/2 = \epsilon \ll 1,
\eea
and we can perform a rotation to go to the basis where the doublets contain the neutral scalar mass eigenstates
\bea
\Phi_h' = \cos\epsilon \Phi_h - \sin\epsilon \Phi_H = \left(
\begin{array}{cc}
 \cos\epsilon G^+ - \sin\epsilon H^+  \\
\frac{1}{\sqrt{2}} \left( v \cos\epsilon + h_{\text{\sc sm}} + \cos\epsilon G^0 - \sin\epsilon A \right) \\
\end{array}
\right),\label{eq:nalignh2}\\
\Phi_H' = \sin\epsilon \Phi_h + \cos\epsilon \Phi_H = \left(
\begin{array}{cc}
 \cos\epsilon H^+ + \sin\epsilon G^+\\
\frac{1}{\sqrt{2}} \left( v\sin\epsilon + H_\perp + \cos\epsilon A + \sin\epsilon G^0 \right) \\
\end{array}
\right),\label{eq:nalignH2}
\eea
where $\epsilon$ is such that\footnote{The lower limit comes from the requirement of perturbativity for the charm quark Yukawa coupling for the new doublet.}
\bea
1\GeV \lesssim v\sin\epsilon \lesssim 100 \GeV \rightarrow 2.5 \lesssim \tan\left(\beta-\alpha\right) \lesssim 250.
\eea
We can then assume the Lagrangian has either an additional global $U(1)$ or $Z_4$ symmetry, with the charges listed in Tab.~\ref{tab:globalsym} (the real scalar needs to be promoted to a complex scalar to give it a $U(1)$ charge). With such symmetries SM fermions are forced to couple only to one of the two doublets of Eq.~(\ref{eq:nalignh2}), (\ref{eq:nalignH2}).
\begin{table}[tb]\centering
\begin{tabular}{|c|cc|}
\hline
Particle & $U(1)$ & $Z_4$
\\ \hline
$\Phi_h'$ & 0 & 0 \\
$\Phi_H'$ & 1 & 1 \\
$S$ & 1 & 1 \\
$Q_L^3$ & $m+3$ & $m-1$ \\
$Q_L^{1,2}$ & $m-1$ & $m$ \\
$u_R^3$ & $m+3$ & $m-1$ \\
$u_R^{1,2}$ & $m+1$ & $m+1$ \\
$d_R^3$ & $m+3$ & $m-1$ \\
$d_R^{1,2}$ & $m-1$ & $m-1$ \\
$\chi_L$ & 1 & 1 \\
$\chi_R$ & 0 & 0 \\
\hline
\end{tabular}
\caption{Example of choice of charges for the global symmetry. Here $m \in \mathbb{Z}$.}\label{tab:globalsym}
\end{table}
We can use these symmetries to discard some terms in the scalar potential, even though we may want to retain soft breaking terms. 
With the $Z_4$ symmetry, we can see that this model is equivalent to up-type BGL models \citep{Branco:1996bq}. Thus FCNC are absent at tree level in the up sector, and are only present in the down sector, suppressed by CKM matrix entries.
The patterns of the $U_{u|L,R}$ matrices that diagonalise the Yukawa matrices are, in the case of the $Z_4$ symmetry, are
\bea
U_{u|L,R} &=& \left(
\begin{array}{ccc}
\times  & \times & 0 \\
\times & \times & 0 \\
0 & 0 & 1\\
\end{array}
\right),
\eea
while the patterns of the Yukawa matrices are
\be
Y_1^U = \left(
\begin{array}{ccc}
0 & 0 & 0 \\
0 & 0 & 0 \\
0 & 0 & \times\\
\end{array}
\right),
Y_2^U = \left(
\begin{array}{ccc}
\times & \times & 0 \\
\times & \times & 0 \\
0 & 0 & 0\\
\end{array}
\right),
Y_1^D = \left(
\begin{array}{ccc}
0 & 0 & 0 \\
0 & 0 & 0 \\
\times & \times & \times\\
\end{array}
\right),
Y_2^D = \left(
\begin{array}{ccc}
\times & \times & \times \\
\times & \times & \times \\
0 & 0 & 0\\
\end{array}
\right).
\ee
In this way, we can choose not only $y_t$, but also $y_c$ to be of order one, and thus the smallness of the lighter quark masses could be explained not by a tiny value of their Yukawa coupling, but rather by the smallness of $v_2$. In this context, the coupling of the charm quark to the 125 GeV Higgs boson would result in being much smaller than the SM, but this is compatible with current experimental results \citep{Bishara:2016jga}.\\
Moreover, in \citep{Botella:2016krk} there is a claim of the presence of fine-tuning in the CKM matrix, given the smallness of $|V_{13}|^2+|V_{23}|^2$. This framework complied with one of the possible explanations of this alignment present in the CKM matrix proposed in \citep{Botella:2016krk} -- however the magnitude of FCNC both at tree and loop level of this model should be analysed and may provide additional constrains on the allowed values of $\epsilon$.

The BGL model induces such FCNC at tree level with couplings for the second doublet given by\footnote{We neglect CP violation also here.} the off-diagonal elements of 
\bea
\frac{N_d+N_d^\dagger}{2v},
\eea
which are
\bea
\frac{m_b}{v \sin 2\epsilon} \left(
\begin{array}{ccc}
0 & 0 & V_{31}^* \\
0 & 0 & V_{32}^* \\
V_{31} & V_{32} & 0\\
\end{array}
\right),
\eea
where $N_d$ is the matrix defined in \cite{Branco:2011iw}:
\begin{align*}
N_d = \frac{1}{\sqrt{2}} U^\dagger_{dL} (v_2 Y^D_1 - v_1 Y^D_2) U_{dR}.
\end{align*}
Using the limits from \citep{Crivellin:2013wna} we get that\footnote{Note that here we are considering the most favourable scenario, where $h_{\text{\sc sm}}$ is a mass eigenstate, and $H_\perp$ is mixing with the singlet, or, in other words, that $S$, written as linear combination of mass eigenstates, does not contain the $125\GeV$ Higgs boson. In general, one will need to consider a full $3\times 3$ mixing for the 3 scalars.}
\bea
\sin 2\epsilon > 0.56 \times 500\GeV\sqrt{\frac{\cos^2\theta}{M_{S_1}^2}+\frac{\sin^2\theta}{M_{S_2}^2}}.
\eea 
For $M_{S_1} = 500\GeV, \theta=\pi/4$ and $M_{S_2}\gg M_{S_1}$, one has
\bea
\sin\epsilon > 0.2, \quad \cos\epsilon<0.98, \quad \cot\epsilon = \tan(\beta-\alpha) < 4.8.
\eea


\newpage

\label{Bibliography}

\lhead{\emph{Bibliography}} 

\bibliography{Bibliography} 

\providecommand{\href}[2]{#2}\begingroup\raggedright\begin{thebibliography}{10}

\bibitem{Busoni:2013lha}
G.~Busoni, A.~De~Simone, E.~Morgante, and A.~Riotto, ``{On the Validity of the
  Effective Field Theory for Dark Matter Searches at the LHC},''
  \href{http://dx.doi.org/10.1016/j.physletb.2013.11.069}{{\em Phys. Lett.}
  {\bfseries B728} (2014) 412--421},
\href{http://arxiv.org/abs/1307.2253}{{\ttfamily arXiv:1307.2253 [hep-ph]}}.

\bibitem{Busoni:2014sya}
G.~Busoni, A.~De~Simone, J.~Gramling, E.~Morgante, and A.~Riotto, ``{On the
  Validity of the Effective Field Theory for Dark Matter Searches at the LHC,
  Part II: Complete Analysis for the $s$-channel},''
  \href{http://dx.doi.org/10.1088/1475-7516/2014/06/060}{{\em JCAP} {\bfseries
  1406} (2014) 060},
\href{http://arxiv.org/abs/1402.1275}{{\ttfamily arXiv:1402.1275 [hep-ph]}}.

\bibitem{Busoni:2014haa}
G.~Busoni, A.~De~Simone, T.~Jacques, E.~Morgante, and A.~Riotto, ``{On the
  Validity of the Effective Field Theory for Dark Matter Searches at the LHC
  Part III: Analysis for the $t$-channel},''
  \href{http://dx.doi.org/10.1088/1475-7516/2014/09/022}{{\em JCAP} {\bfseries
  1409} (2014) 022},
\href{http://arxiv.org/abs/1405.3101}{{\ttfamily arXiv:1405.3101 [hep-ph]}}.

\bibitem{Buchmueller:2013dya}
O.~Buchmueller, M.~J. Dolan, and C.~McCabe, ``{Beyond Effective Field Theory
  for Dark Matter Searches at the LHC},''
  \href{http://dx.doi.org/10.1007/JHEP01(2014)025}{{\em JHEP} {\bfseries 01}
  (2014) 025},
\href{http://arxiv.org/abs/1308.6799}{{\ttfamily arXiv:1308.6799 [hep-ph]}}.

\bibitem{Abdallah:2014hon}
J.~Abdallah {\em et~al.}, ``{Simplified Models for Dark Matter and Missing
  Energy Searches at the LHC},''
\href{http://arxiv.org/abs/1409.2893}{{\ttfamily arXiv:1409.2893 [hep-ph]}}.

\bibitem{Shoemaker:2011vi}
I.~M. Shoemaker and L.~Vecchi, ``{Unitarity and Monojet Bounds on Models for
  DAMA, CoGeNT, and CRESST-II},''
  \href{http://dx.doi.org/10.1103/PhysRevD.86.015023}{{\em Phys. Rev.}
  {\bfseries D86} (2012) 015023},
\href{http://arxiv.org/abs/1112.5457}{{\ttfamily arXiv:1112.5457 [hep-ph]}}.

\bibitem{Abdallah:2015ter}
J.~Abdallah {\em et~al.}, ``{Simplified Models for Dark Matter Searches at the
  LHC},'' \href{http://dx.doi.org/10.1016/j.dark.2015.08.001}{{\em Phys. Dark
  Univ.} {\bfseries 9-10} (2015) 8--23},
\href{http://arxiv.org/abs/1506.03116}{{\ttfamily arXiv:1506.03116 [hep-ph]}}.

\bibitem{Abercrombie:2015wmb}
D.~Abercrombie {\em et~al.}, ``{Dark Matter Benchmark Models for Early LHC
  Run-2 Searches: Report of the ATLAS/CMS Dark Matter Forum},''
\href{http://arxiv.org/abs/1507.00966}{{\ttfamily arXiv:1507.00966 [hep-ex]}}.

\bibitem{Boveia:2016mrp}
G.~Busoni {\em et~al.}, ``{Recommendations on presenting LHC searches for
  missing transverse energy signals using simplified $s$-channel models of dark
  matter},''
\href{http://arxiv.org/abs/1603.04156}{{\ttfamily arXiv:1603.04156 [hep-ex]}}.

\bibitem{Jacques:2016dqz}
T.~Jacques, A.~Katz, E.~Morgante, D.~Racco, M.~Rameez, and A.~Riotto,
  ``{Complementarity of DM Searches in a Consistent Simplified Model: the Case
  of Z'},''
\href{http://arxiv.org/abs/1605.06513}{{\ttfamily arXiv:1605.06513 [hep-ph]}}.

\bibitem{Buckley:2014fba}
M.~R. Buckley, D.~Feld, and D.~Goncalves, ``{Scalar Simplified Models for Dark
  Matter},'' \href{http://dx.doi.org/10.1103/PhysRevD.91.015017}{{\em Phys.
  Rev.} {\bfseries D91} (2015) 015017},
\href{http://arxiv.org/abs/1410.6497}{{\ttfamily arXiv:1410.6497 [hep-ph]}}.

\bibitem{Harris:2014hga}
P.~Harris, V.~V. Khoze, M.~Spannowsky, and C.~Williams, ``{Constraining Dark
  Sectors at Colliders: Beyond the Effective Theory Approach},''
  \href{http://dx.doi.org/10.1103/PhysRevD.91.055009}{{\em Phys. Rev.}
  {\bfseries D91} (2015) 055009},
\href{http://arxiv.org/abs/1411.0535}{{\ttfamily arXiv:1411.0535 [hep-ph]}}.

\bibitem{Bell:2015sza}
N.~F. Bell, Y.~Cai, J.~B. Dent, R.~K. Leane, and T.~J. Weiler, ``{Dark matter
  at the LHC: Effective field theories and gauge invariance},''
  \href{http://dx.doi.org/10.1103/PhysRevD.92.053008}{{\em Phys. Rev.}
  {\bfseries D92} no.~5, (2015) 053008},
\href{http://arxiv.org/abs/1503.07874}{{\ttfamily arXiv:1503.07874 [hep-ph]}}.

\bibitem{Bell:2015rdw}
N.~F. Bell, Y.~Cai, and R.~K. Leane, ``{Mono-W Dark Matter Signals at the LHC:
  Simplified Model Analysis},''
  \href{http://dx.doi.org/10.1088/1475-7516/2016/01/051}{{\em JCAP} {\bfseries
  1601} no.~01, (2016) 051},
\href{http://arxiv.org/abs/1512.00476}{{\ttfamily arXiv:1512.00476 [hep-ph]}}.

\bibitem{Haisch:2016usn}
U.~Haisch, F.~Kahlhoefer, and T.~M.~P. Tait, ``{On Mono-W Signatures in Spin-1
  Simplified Models},''
  \href{http://dx.doi.org/10.1016/j.physletb.2016.06.063}{{\em Phys. Lett.}
  {\bfseries B760} (2016) 207--213},
\href{http://arxiv.org/abs/1603.01267}{{\ttfamily arXiv:1603.01267 [hep-ph]}}.

\bibitem{Englert:2016joy}
C.~Englert, M.~McCullough, and M.~Spannowsky, ``{S-Channel Dark Matter
  Simplified Models and Unitarity},''
\href{http://arxiv.org/abs/1604.07975}{{\ttfamily arXiv:1604.07975 [hep-ph]}}.

\bibitem{Kahlhoefer:2015bea}
F.~Kahlhoefer, K.~Schmidt-Hoberg, T.~Schwetz, and S.~Vogl, ``{Implications of
  unitarity and gauge invariance for simplified dark matter models},''
  \href{http://dx.doi.org/10.1007/JHEP02(2016)016}{{\em JHEP} {\bfseries 02}
  (2016) 016},
\href{http://arxiv.org/abs/1510.02110}{{\ttfamily arXiv:1510.02110 [hep-ph]}}.

\bibitem{Bell:2016fqf}
N.~F. Bell, Y.~Cai, and R.~K. Leane, ``{Dark Forces in the Sky: Signals from Z'
  and the Dark Higgs},''
  \href{http://dx.doi.org/10.1088/1475-7516/2016/08/001}{{\em JCAP} {\bfseries
  1608} no.~08, (2016) 001},
\href{http://arxiv.org/abs/1605.09382}{{\ttfamily arXiv:1605.09382 [hep-ph]}}.

\bibitem{Ko:2016zxg}
P.~Ko, A.~Natale, M.~Park, and H.~Yokoya, ``{Simplified DM models with the full
  SM gauge symmetry : the case of $t$-channel colored scalar mediators},''
\href{http://arxiv.org/abs/1605.07058}{{\ttfamily arXiv:1605.07058 [hep-ph]}}.

\bibitem{Duerr:2016tmh}
M.~Duerr, F.~Kahlhoefer, K.~Schmidt-Hoberg, T.~Schwetz, and S.~Vogl, ``{How to
  save the WIMP: global analysis of a dark matter model with two s-channel
  mediators},'' \href{http://dx.doi.org/10.1007/JHEP09(2016)042}{{\em JHEP}
  {\bfseries 09} (2016) 042},
\href{http://arxiv.org/abs/1606.07609}{{\ttfamily arXiv:1606.07609 [hep-ph]}}.

\bibitem{Bell:2016uhg}
N.~F. Bell, Y.~Cai, and R.~K. Leane, ``{Impact of Mass Generation for
  Simplified Dark Matter Models},''
\href{http://arxiv.org/abs/1610.03063}{{\ttfamily arXiv:1610.03063 [hep-ph]}}.

\bibitem{Khoze:2015sra}
V.~V. Khoze, G.~Ro, and M.~Spannowsky, ``{Spectroscopy of scalar mediators to
  dark matter at the LHC and at 100 TeV},''
  \href{http://dx.doi.org/10.1103/PhysRevD.92.075006}{{\em Phys. Rev.}
  {\bfseries D92} no.~7, (2015) 075006},
\href{http://arxiv.org/abs/1505.03019}{{\ttfamily arXiv:1505.03019 [hep-ph]}}.

\bibitem{Baek:2015lna}
S.~Baek, P.~Ko, M.~Park, W.-I. Park, and C.~Yu, ``{Beyond the Dark matter
  effective field theory and a simplified model approach at colliders},''
  \href{http://dx.doi.org/10.1016/j.physletb.2016.03.026}{{\em Phys. Lett.}
  {\bfseries B756} (2016) 289--294},
\href{http://arxiv.org/abs/1506.06556}{{\ttfamily arXiv:1506.06556 [hep-ph]}}.

\bibitem{Bauer:2016gys}
M.~Bauer {\em et~al.}, ``{Towards the next generation of simplified Dark Matter
  models},''
\href{http://arxiv.org/abs/1607.06680}{{\ttfamily arXiv:1607.06680 [hep-ex]}}.

\bibitem{Robens:2016xkb}
T.~Robens and T.~Stefaniak, ``{LHC Benchmark Scenarios for the Real Higgs
  Singlet Extension of the Standard Model},''
  \href{http://dx.doi.org/10.1140/epjc/s10052-016-4115-8}{{\em Eur. Phys. J.}
  {\bfseries C76} no.~5, (2016) 268},
\href{http://arxiv.org/abs/1601.07880}{{\ttfamily arXiv:1601.07880 [hep-ph]}}.

\bibitem{Wang:2015cda}
Z.-W. Wang, T.~G. Steele, T.~Hanif, and R.~B. Mann, ``{Conformal Complex
  Singlet Extension of the Standard Model: Scenario for Dark Matter and a
  Second Higgs Boson},'' \href{http://dx.doi.org/10.1007/JHEP08(2016)065}{{\em
  JHEP} {\bfseries 08} (2016) 065},
\href{http://arxiv.org/abs/1510.04321}{{\ttfamily arXiv:1510.04321 [hep-ph]}}.

\bibitem{Lopez-Val:2014jva}
D.~López-Val and T.~Robens, ``{Δr and the W-boson mass in the singlet
  extension of the standard model},''
  \href{http://dx.doi.org/10.1103/PhysRevD.90.114018}{{\em Phys. Rev.}
  {\bfseries D90} (2014) 114018},
\href{http://arxiv.org/abs/1406.1043}{{\ttfamily arXiv:1406.1043 [hep-ph]}}.

\bibitem{Costa:2015llh}
R.~Costa, M.~Mühlleitner, M.~O.~P. Sampaio, and R.~Santos, ``{Singlet
  Extensions of the Standard Model at LHC Run 2: Benchmarks and Comparison with
  the NMSSM},'' \href{http://dx.doi.org/10.1007/JHEP06(2016)034}{{\em JHEP}
  {\bfseries 06} (2016) 034},
\href{http://arxiv.org/abs/1512.05355}{{\ttfamily arXiv:1512.05355 [hep-ph]}}.

\bibitem{Dupuis:2016fda}
G.~Dupuis, ``{Collider Constraints and Prospects of a Scalar Singlet Extension
  to Higgs Portal Dark Matter},''
  \href{http://dx.doi.org/10.1007/JHEP07(2016)008}{{\em JHEP} {\bfseries 07}
  (2016) 008},
\href{http://arxiv.org/abs/1604.04552}{{\ttfamily arXiv:1604.04552 [hep-ph]}}.

\bibitem{Balazs:2016tbi}
C.~Balazs, A.~Fowlie, A.~Mazumdar, and G.~White, ``{Gravitational waves at
  aLIGO and vacuum stability with a scalar singlet extension of the Standard
  Model},''
\href{http://arxiv.org/abs/1611.01617}{{\ttfamily arXiv:1611.01617 [hep-ph]}}.

\bibitem{Branco:2011iw}
G.~C. Branco, P.~M. Ferreira, L.~Lavoura, M.~N. Rebelo, M.~Sher, and J.~P.
  Silva, ``{Theory and phenomenology of two-Higgs-doublet models},''
  \href{http://dx.doi.org/10.1016/j.physrep.2012.02.002}{{\em Phys. Rept.}
  {\bfseries 516} (2012) 1--102},
\href{http://arxiv.org/abs/1106.0034}{{\ttfamily arXiv:1106.0034 [hep-ph]}}.

\bibitem{Amaldi:1991cn}
U.~Amaldi, W.~de~Boer, and H.~Furstenau, ``{Comparison of grand unified
  theories with electroweak and strong coupling constants measured at LEP},''
\href{http://dx.doi.org/10.1016/0370-2693(91)91641-8}{{\em Phys. Lett.}
  {\bfseries B260} (1991) 447--455}.

\bibitem{Carena:1995wu}
M.~Carena, M.~Quiros, and C.~E.~M. Wagner, ``{Effective potential methods and
  the Higgs mass spectrum in the MSSM},''
  \href{http://dx.doi.org/10.1016/0550-3213(95)00665-6}{{\em Nucl. Phys.}
  {\bfseries B461} (1996) 407--436},
\href{http://arxiv.org/abs/hep-ph/9508343}{{\ttfamily arXiv:hep-ph/9508343
  [hep-ph]}}.

\bibitem{Bhattacharyya:2015nca}
G.~Bhattacharyya and D.~Das, ``{Scalar sector of two-Higgs-doublet models: A
  minireview},'' \href{http://dx.doi.org/10.1007/s12043-016-1252-4}{{\em
  Pramana} {\bfseries 87} no.~3, (2016) 40},
\href{http://arxiv.org/abs/1507.06424}{{\ttfamily arXiv:1507.06424 [hep-ph]}}.

\bibitem{Fayet:1976et}
P.~Fayet, ``{Supersymmetry and Weak, Electromagnetic and Strong
  Interactions},''
\href{http://dx.doi.org/10.1016/0370-2693(76)90319-1}{{\em Phys. Lett.}
  {\bfseries B64} (1976) 159}.

\bibitem{Gunion:1984yn}
J.~F. Gunion and H.~E. Haber, ``{Higgs Bosons in Supersymmetric Models. 1.},''
  \href{http://dx.doi.org/10.1016/0550-3213(86)90340-8,
  10.1016/0550-3213(93)90653-7}{{\em Nucl. Phys.} {\bfseries B272} (1986) 1}.
[Erratum: Nucl. Phys.B402,567(1993)].

\bibitem{King:2012tr}
S.~F. King, M.~Mühlleitner, R.~Nevzorov, and K.~Walz, ``{Natural NMSSM Higgs
  Bosons},'' \href{http://dx.doi.org/10.1016/j.nuclphysb.2013.01.020}{{\em
  Nucl. Phys.} {\bfseries B870} (2013) 323--352},
\href{http://arxiv.org/abs/1211.5074}{{\ttfamily arXiv:1211.5074 [hep-ph]}}.

\bibitem{Chen:2013jvg}
C.-Y. Chen, M.~Freid, and M.~Sher, ``{Next-to-minimal two Higgs doublet
  model},'' \href{http://dx.doi.org/10.1103/PhysRevD.89.075009}{{\em Phys.
  Rev.} {\bfseries D89} no.~7, (2014) 075009},
\href{http://arxiv.org/abs/1312.3949}{{\ttfamily arXiv:1312.3949 [hep-ph]}}.

\bibitem{Kanemura:2015fra}
S.~Kanemura, M.~Kikuchi, and K.~Yagyu, ``{Radiative corrections to the Higgs
  boson couplings in the model with an additional real singlet scalar field},''
  \href{http://dx.doi.org/10.1016/j.nuclphysb.2016.04.005}{{\em Nucl. Phys.}
  {\bfseries B907} (2016) 286--322},
\href{http://arxiv.org/abs/1511.06211}{{\ttfamily arXiv:1511.06211 [hep-ph]}}.

\bibitem{vonBuddenbrock:2016rmr}
S.~von Buddenbrock, N.~Chakrabarty, A.~S. Cornell, D.~Kar, M.~Kumar, T.~Mandal,
  B.~Mellado, B.~Mukhopadhyaya, R.~G. Reed, and X.~Ruan, ``{Phenomenological
  signatures of additional scalar bosons at the LHC},''
  \href{http://dx.doi.org/10.1140/epjc/s10052-016-4435-8}{{\em Eur. Phys. J.}
  {\bfseries C76} no.~10, (2016) 580},
\href{http://arxiv.org/abs/1606.01674}{{\ttfamily arXiv:1606.01674 [hep-ph]}}.

\bibitem{Ko:2016ybp}
P.~Ko and J.~Li, ``{Interference effects of two scalar boson propagators on the
  LHC search for the singlet fermion DM},''
\href{http://arxiv.org/abs/1610.03997}{{\ttfamily arXiv:1610.03997 [hep-ph]}}.

\bibitem{Kumar:2013iva}
J.~Kumar and D.~Marfatia, ``{Matrix element analyses of dark matter scattering
  and annihilation},'' \href{http://dx.doi.org/10.1103/PhysRevD.88.014035}{{\em
  Phys. Rev.} {\bfseries D88} no.~1, (2013) 014035},
\href{http://arxiv.org/abs/1305.1611}{{\ttfamily arXiv:1305.1611 [hep-ph]}}.

\bibitem{Ipek:2014gua}
S.~Ipek, D.~McKeen, and A.~E. Nelson, ``{A Renormalizable Model for the
  Galactic Center Gamma Ray Excess from Dark Matter Annihilation},''
  \href{http://dx.doi.org/10.1103/PhysRevD.90.055021}{{\em Phys. Rev.}
  {\bfseries D90} no.~5, (2014) 055021},
\href{http://arxiv.org/abs/1404.3716}{{\ttfamily arXiv:1404.3716 [hep-ph]}}.

\bibitem{Berlin:2015wwa}
A.~Berlin, S.~Gori, T.~Lin, and L.-T. Wang, ``{Pseudoscalar Portal Dark
  Matter},'' \href{http://dx.doi.org/10.1103/PhysRevD.92.015005}{{\em Phys.
  Rev.} {\bfseries D92} (2015) 015005},
\href{http://arxiv.org/abs/1502.06000}{{\ttfamily arXiv:1502.06000 [hep-ph]}}.

\bibitem{Goncalves:2016iyg}
D.~Goncalves, P.~A.~N. Machado, and J.~M. No, ``{Simplified Models for Dark
  Matter Face their Consistent Completions},''
\href{http://arxiv.org/abs/1611.04593}{{\ttfamily arXiv:1611.04593 [hep-ph]}}.

\bibitem{No:2015xqa}
J.~M. No, ``{Looking through the pseudoscalar portal into dark matter: Novel
  mono-Higgs and mono-Z signatures at the LHC},''
  \href{http://dx.doi.org/10.1103/PhysRevD.93.031701}{{\em Phys. Rev.}
  {\bfseries D93} no.~3, (2016) 031701},
\href{http://arxiv.org/abs/1509.01110}{{\ttfamily arXiv:1509.01110 [hep-ph]}}.

\bibitem{Bauer:2017ota}
M.~Bauer, U.~Haisch, and F.~Kahlhoefer, ``{Simplified dark matter models with
  two Higgs doublets: I. Pseudoscalar mediators},''
\href{http://arxiv.org/abs/1701.07427}{{\ttfamily arXiv:1701.07427 [hep-ph]}}.

\bibitem{Haisch:2016gry}
U.~Haisch, P.~Pani, and G.~Polesello, ``{Determining the CP nature of spin-0
  mediators in associated production of dark matter and $t \bar t$ pairs},''
\href{http://arxiv.org/abs/1611.09841}{{\ttfamily arXiv:1611.09841 [hep-ph]}}.

\bibitem{Aad:2015uga}
{\bfseries ATLAS} Collaboration, G.~Aad {\em et~al.}, ``{Search for invisible
  decays of the Higgs boson produced in association with a hadronically
  decaying vector boson in $pp$ collisions at $\sqrt{s}$ = 8 TeV with the ATLAS
  detector},'' \href{http://dx.doi.org/10.1140/epjc/s10052-015-3551-1}{{\em
  Eur. Phys. J.} {\bfseries C75} no.~7, (2015) 337},
\href{http://arxiv.org/abs/1504.04324}{{\ttfamily arXiv:1504.04324 [hep-ex]}}.

\bibitem{Chatrchyan:2014tja}
{\bfseries CMS} Collaboration, S.~Chatrchyan {\em et~al.}, ``{Search for
  invisible decays of Higgs bosons in the vector boson fusion and associated ZH
  production modes},''
  \href{http://dx.doi.org/10.1140/epjc/s10052-014-2980-6}{{\em Eur. Phys. J.}
  {\bfseries C74} (2014) 2980},
\href{http://arxiv.org/abs/1404.1344}{{\ttfamily arXiv:1404.1344 [hep-ex]}}.

\bibitem{ATLAS-CONF-2015-044}
``{Measurements of the Higgs boson production and decay rates and constraints
  on its couplings from a combined ATLAS and CMS analysis of the LHC pp
  collision data at $\sqrt{s}$ = 7 and 8 TeV},'' Tech. Rep.
  ATLAS-CONF-2015-044, CERN, Geneva, Sep, 2015.
\newblock \url{http://cds.cern.ch/record/2052552}.

\bibitem{CMS:2016uxr}
{\bfseries CMS} Collaboration, C.~Collaboration,
``{Search for Dark Matter produced in association with bottom quarks},''.

\bibitem{CMS:2016pod}
{\bfseries CMS} Collaboration, C.~Collaboration,
``{Search for dark matter in final states with an energetic jet, or a
  hadronically decaying W or Z boson using $12.9~\mathrm{fb}^{-1}$ of data at
  $\sqrt{s} = 13~\mathrm{TeV}$},''.

\bibitem{CMS:2016mxc}
{\bfseries CMS} Collaboration, C.~Collaboration,
``{Search for dark matter in association with a top quark pair at sqrt(s)=13
  TeV},''.

\bibitem{ATLAS-CONF-2016-086}
{\bfseries ATLAS Collaboration} Collaboration, ``{Search for Dark Matter
  production associated with bottom quarks with 13.3 fb−1 of pp collisions at
  √s = 13 TeV with the ATLAS detector at the LHC},'' Tech. Rep.
  ATLAS-CONF-2016-086, CERN, Geneva, Aug, 2016.
\newblock \url{http://cds.cern.ch/record/2206279}.

\bibitem{ATLAS-CONF-2016-077}
{\bfseries ATLAS Collaboration} Collaboration, ``{Search for the Supersymmetric
  Partner of the Top Quark in the Jets+Emiss Final State at sqrt(s) = 13
  TeV},'' Tech. Rep. ATLAS-CONF-2016-077, CERN, Geneva, Aug, 2016.
\newblock \url{http://cds.cern.ch/record/2206250}.

\bibitem{Gunion:2002zf}
J.~F. Gunion and H.~E. Haber, ``{The CP conserving two Higgs doublet model: The
  Approach to the decoupling limit},''
  \href{http://dx.doi.org/10.1103/PhysRevD.67.075019}{{\em Phys. Rev.}
  {\bfseries D67} (2003) 075019},
\href{http://arxiv.org/abs/hep-ph/0207010}{{\ttfamily arXiv:hep-ph/0207010
  [hep-ph]}}.

\bibitem{Carena:2013ooa}
M.~Carena, I.~Low, N.~R. Shah, and C.~E.~M. Wagner, ``{Impersonating the
  Standard Model Higgs Boson: Alignment without Decoupling},''
  \href{http://dx.doi.org/10.1007/JHEP04(2014)015}{{\em JHEP} {\bfseries 04}
  (2014) 015},
\href{http://arxiv.org/abs/1310.2248}{{\ttfamily arXiv:1310.2248 [hep-ph]}}.

\bibitem{Dev:2014yca}
P.~S. Bhupal~Dev and A.~Pilaftsis, ``{Maximally Symmetric Two Higgs Doublet
  Model with Natural Standard Model Alignment},''
  \href{http://dx.doi.org/10.1007/JHEP11(2015)147,
  10.1007/JHEP12(2014)024}{{\em JHEP} {\bfseries 12} (2014) 024},
  \href{http://arxiv.org/abs/1408.3405}{{\ttfamily arXiv:1408.3405 [hep-ph]}}.
[Erratum: JHEP11,147(2015)].

\bibitem{Dev:2015bta}
P.~S.~B. Dev and A.~Pilaftsis, ``{Natural Standard Model Alignment in the Two
  Higgs Doublet Model},''
  \href{http://dx.doi.org/10.1088/1742-6596/631/1/012030}{{\em J. Phys. Conf.
  Ser.} {\bfseries 631} no.~1, (2015) 012030},
\href{http://arxiv.org/abs/1503.09140}{{\ttfamily arXiv:1503.09140 [hep-ph]}}.

\bibitem{Pilaftsis:2016erj}
A.~Pilaftsis, ``{Symmetries for standard model alignment in multi-Higgs doublet
  models},'' \href{http://dx.doi.org/10.1103/PhysRevD.93.075012}{{\em Phys.
  Rev.} {\bfseries D93} no.~7, (2016) 075012},
\href{http://arxiv.org/abs/1602.02017}{{\ttfamily arXiv:1602.02017 [hep-ph]}}.

\bibitem{Draper:2016cag}
P.~Draper, H.~E. Haber, and J.~T. Ruderman, ``{Partially Natural Two Higgs
  Doublet Models},'' \href{http://dx.doi.org/10.1007/JHEP06(2016)124}{{\em
  JHEP} {\bfseries 06} (2016) 124},
\href{http://arxiv.org/abs/1605.03237}{{\ttfamily arXiv:1605.03237 [hep-ph]}}.

\bibitem{D'Ambrosio:2002ex}
G.~D'Ambrosio, G.~F. Giudice, G.~Isidori, and A.~Strumia, ``{Minimal flavor
  violation: An Effective field theory approach},''
  \href{http://dx.doi.org/10.1016/S0550-3213(02)00836-2}{{\em Nucl. Phys.}
  {\bfseries B645} (2002) 155--187},
\href{http://arxiv.org/abs/hep-ph/0207036}{{\ttfamily arXiv:hep-ph/0207036
  [hep-ph]}}.

\bibitem{Buras:2010mh}
A.~J. Buras, M.~V. Carlucci, S.~Gori, and G.~Isidori, ``{Higgs-mediated FCNCs:
  Natural Flavour Conservation vs. Minimal Flavour Violation},''
  \href{http://dx.doi.org/10.1007/JHEP10(2010)009}{{\em JHEP} {\bfseries 10}
  (2010) 009},
\href{http://arxiv.org/abs/1005.5310}{{\ttfamily arXiv:1005.5310 [hep-ph]}}.

\bibitem{Pich:2009sp}
A.~Pich and P.~Tuzon, ``{Yukawa Alignment in the Two-Higgs-Doublet Model},''
  \href{http://dx.doi.org/10.1103/PhysRevD.80.091702}{{\em Phys. Rev.}
  {\bfseries D80} (2009) 091702},
\href{http://arxiv.org/abs/0908.1554}{{\ttfamily arXiv:0908.1554 [hep-ph]}}.

\bibitem{Tuzon:2010vt}
P.~Tuzon and A.~Pich, ``{The Aligned two-Higgs Doublet model},'' {\em Acta
  Phys. Polon. Supp.} {\bfseries 3} (2010) 215--220,
\href{http://arxiv.org/abs/1001.0293}{{\ttfamily arXiv:1001.0293 [hep-ph]}}.

\bibitem{Pich:2010ic}
A.~Pich, ``{Flavour constraints on multi-Higgs-doublet models: Yukawa
  alignment},'' \href{http://dx.doi.org/10.1016/j.nuclphysbps.2010.12.030}{{\em
  Nucl. Phys. Proc. Suppl.} {\bfseries 209} (2010) 182--187},
\href{http://arxiv.org/abs/1010.5217}{{\ttfamily arXiv:1010.5217 [hep-ph]}}.

\bibitem{Braeuninger:2010td}
C.~B. Braeuninger, A.~Ibarra, and C.~Simonetto, ``{Radiatively induced flavour
  violation in the general two-Higgs doublet model with Yukawa alignment},''
  \href{http://dx.doi.org/10.1016/j.physletb.2010.07.039}{{\em Phys. Lett.}
  {\bfseries B692} (2010) 189--195},
\href{http://arxiv.org/abs/1005.5706}{{\ttfamily arXiv:1005.5706 [hep-ph]}}.

\bibitem{Botella:2015yfa}
F.~J. Botella, G.~C. Branco, A.~M. Coutinho, M.~N. Rebelo, and J.~I.
  Silva-Marcos, ``{Natural Quasi-Alignment with two Higgs Doublets and RGE
  Stability},'' \href{http://dx.doi.org/10.1140/epjc/s10052-015-3487-5}{{\em
  Eur. Phys. J.} {\bfseries C75} (2015) 286},
\href{http://arxiv.org/abs/1501.07435}{{\ttfamily arXiv:1501.07435 [hep-ph]}}.

\bibitem{Jung:2010ik}
M.~Jung, A.~Pich, and P.~Tuzon, ``{Charged-Higgs phenomenology in the Aligned
  two-Higgs-doublet model},''
  \href{http://dx.doi.org/10.1007/JHEP11(2010)003}{{\em JHEP} {\bfseries 11}
  (2010) 003},
\href{http://arxiv.org/abs/1006.0470}{{\ttfamily arXiv:1006.0470 [hep-ph]}}.

\bibitem{Enomoto:2015wbn}
T.~Enomoto and R.~Watanabe, ``{Flavor constraints on the Two Higgs Doublet
  Models of Z$_{2}$ symmetric and aligned types},''
  \href{http://dx.doi.org/10.1007/JHEP05(2016)002}{{\em JHEP} {\bfseries 05}
  (2016) 002},
\href{http://arxiv.org/abs/1511.05066}{{\ttfamily arXiv:1511.05066 [hep-ph]}}.

\bibitem{Evans:2011wj}
J.~L. Evans, B.~Feldstein, W.~Klemm, H.~Murayama, and T.~T. Yanagida,
  ``{Hermitian Flavor Violation},''
  \href{http://dx.doi.org/10.1016/j.physletb.2011.08.059}{{\em Phys. Lett.}
  {\bfseries B703} (2011) 599--605},
\href{http://arxiv.org/abs/1106.1734}{{\ttfamily arXiv:1106.1734 [hep-ph]}}.

\bibitem{Altmannshofer:2016zrn}
W.~Altmannshofer, J.~Eby, S.~Gori, M.~Lotito, M.~Martone, and D.~Tuckler,
  ``{Collider Signatures of Flavorful Higgs Bosons},'' {\em Submitted to: Phys.
  Rev. D} (2016) ,
\href{http://arxiv.org/abs/1610.02398}{{\ttfamily arXiv:1610.02398 [hep-ph]}}.

\bibitem{Arroyo:2013tna}
M.~Arroyo, J.~L. Diaz-Cruz, E.~Diaz, and J.~A. Orduz-Ducuara, ``{Flavor
  Violating Higgs signals in the Texturized Two-Higgs Doublet Model
  (2HDM-Tx)},''
\href{http://arxiv.org/abs/1306.2343}{{\ttfamily arXiv:1306.2343 [hep-ph]}}.

\bibitem{Bishara:2015cha}
F.~Bishara, J.~Brod, P.~Uttayarat, and J.~Zupan, ``{Nonstandard Yukawa
  Couplings and Higgs Portal Dark Matter},''
  \href{http://dx.doi.org/10.1007/JHEP01(2016)010}{{\em JHEP} {\bfseries 01}
  (2016) 010},
\href{http://arxiv.org/abs/1504.04022}{{\ttfamily arXiv:1504.04022 [hep-ph]}}.

\bibitem{Cvetic:1998uw}
G.~Cvetic, C.~S. Kim, and S.~S. Hwang, ``{Higgs mediated flavor changing
  neutral currents in the general framework with two Higgs doublets: An RGE
  analysis},'' \href{http://dx.doi.org/10.1103/PhysRevD.58.116003}{{\em Phys.
  Rev.} {\bfseries D58} (1998) 116003},
\href{http://arxiv.org/abs/hep-ph/9806282}{{\ttfamily arXiv:hep-ph/9806282
  [hep-ph]}}.

\bibitem{Desai:2003xz}
B.~R. Desai and A.~R. Vaucher, ``{Solutions to the renormalization group
  equations for Yukawa matrices as an answer to the quark and lepton mass
  problem},''
\href{http://arxiv.org/abs/hep-ph/0309102}{{\ttfamily arXiv:hep-ph/0309102
  [hep-ph]}}.

\bibitem{Kim:2008pp}
Y.~G. Kim, K.~Y. Lee, and S.~Shin, ``{Singlet fermionic dark matter},''
  \href{http://dx.doi.org/10.1088/1126-6708/2008/05/100}{{\em JHEP} {\bfseries
  05} (2008) 100},
\href{http://arxiv.org/abs/0803.2932}{{\ttfamily arXiv:0803.2932 [hep-ph]}}.

\bibitem{Baek:2011aa}
S.~Baek, P.~Ko, and W.-I. Park, ``{Search for the Higgs portal to a singlet
  fermionic dark matter at the LHC},''
  \href{http://dx.doi.org/10.1007/JHEP02(2012)047}{{\em JHEP} {\bfseries 02}
  (2012) 047},
\href{http://arxiv.org/abs/1112.1847}{{\ttfamily arXiv:1112.1847 [hep-ph]}}.

\bibitem{LopezHonorez:2012kv}
L.~Lopez-Honorez, T.~Schwetz, and J.~Zupan, ``{Higgs portal, fermionic dark
  matter, and a Standard Model like Higgs at 125 GeV},''
  \href{http://dx.doi.org/10.1016/j.physletb.2012.07.017}{{\em Phys. Lett.}
  {\bfseries B716} (2012) 179--185},
\href{http://arxiv.org/abs/1203.2064}{{\ttfamily arXiv:1203.2064 [hep-ph]}}.

\bibitem{DelNobile:2013sia}
M.~Cirelli, E.~Del~Nobile, and P.~Panci, ``{Tools for model-independent bounds
  in direct dark matter searches},''
  \href{http://dx.doi.org/10.1088/1475-7516/2013/10/019}{{\em JCAP} {\bfseries
  1310} (2013) 019},
\href{http://arxiv.org/abs/1307.5955}{{\ttfamily arXiv:1307.5955 [hep-ph]}}.

\bibitem{Akerib:2013tjd}
{\bfseries LUX} Collaboration, D.~S. Akerib {\em et~al.}, ``{First results from
  the LUX dark matter experiment at the Sanford Underground Research
  Facility},'' \href{http://dx.doi.org/10.1103/PhysRevLett.112.091303}{{\em
  Phys. Rev. Lett.} {\bfseries 112} (2014) 091303},
\href{http://arxiv.org/abs/1310.8214}{{\ttfamily arXiv:1310.8214
  [astro-ph.CO]}}.

\bibitem{Ellis:2016jkw}
J.~Ellis, ``{TikZ-Feynman: Feynman diagrams with TikZ},''
\href{http://arxiv.org/abs/1601.05437}{{\ttfamily arXiv:1601.05437 [hep-ph]}}.

\bibitem{Bell:2016obu}
N.~Bell, G.~Busoni, A.~Kobakhidze, D.~M. Long, and M.~A. Schmidt,
  ``{Unitarisation of EFT Amplitudes for Dark Matter Searches at the LHC},''
  \href{http://dx.doi.org/10.1007/JHEP08(2016)125}{{\em JHEP} {\bfseries 08}
  (2016) 125},
\href{http://arxiv.org/abs/1606.02722}{{\ttfamily arXiv:1606.02722 [hep-ph]}}.

\bibitem{Kanemura:2015ska}
S.~Kanemura and K.~Yagyu, ``{Unitarity bound in the most general two Higgs
  doublet model},''
  \href{http://dx.doi.org/10.1016/j.physletb.2015.10.047}{{\em Phys. Lett.}
  {\bfseries B751} (2015) 289--296},
\href{http://arxiv.org/abs/1509.06060}{{\ttfamily arXiv:1509.06060 [hep-ph]}}.

\bibitem{Lopez-Val:2013yba}
D.~López-Val, T.~Plehn, and M.~Rauch, ``{Measuring extended Higgs sectors as a
  consistent free couplings model},''
  \href{http://dx.doi.org/10.1007/JHEP10(2013)134}{{\em JHEP} {\bfseries 10}
  (2013) 134},
\href{http://arxiv.org/abs/1308.1979}{{\ttfamily arXiv:1308.1979 [hep-ph]}}.

\bibitem{Ferreira:2010xe}
P.~M. Ferreira, L.~Lavoura, and J.~P. Silva, ``{Renormalization-group
  constraints on Yukawa alignment in multi-Higgs-doublet models},''
  \href{http://dx.doi.org/10.1016/j.physletb.2010.04.033}{{\em Phys. Lett.}
  {\bfseries B688} (2010) 341--344},
\href{http://arxiv.org/abs/1001.2561}{{\ttfamily arXiv:1001.2561 [hep-ph]}}.

\bibitem{Crivellin:2013wna}
A.~Crivellin, A.~Kokulu, and C.~Greub, ``{Flavor-phenomenology of
  two-Higgs-doublet models with generic Yukawa structure},''
  \href{http://dx.doi.org/10.1103/PhysRevD.87.094031}{{\em Phys. Rev.}
  {\bfseries D87} no.~9, (2013) 094031},
\href{http://arxiv.org/abs/1303.5877}{{\ttfamily arXiv:1303.5877 [hep-ph]}}.

\bibitem{Branco:1996bq}
G.~C. Branco, W.~Grimus, and L.~Lavoura, ``{Relating the scalar flavor changing
  neutral couplings to the CKM matrix},''
  \href{http://dx.doi.org/10.1016/0370-2693(96)00494-7}{{\em Phys. Lett.}
  {\bfseries B380} (1996) 119--126},
\href{http://arxiv.org/abs/hep-ph/9601383}{{\ttfamily arXiv:hep-ph/9601383
  [hep-ph]}}.

\bibitem{Bishara:2016jga}
F.~Bishara, U.~Haisch, P.~F. Monni, and E.~Re, ``{Constraining Light-Quark
  Yukawa Couplings from Higgs Distributions},''
\href{http://arxiv.org/abs/1606.09253}{{\ttfamily arXiv:1606.09253 [hep-ph]}}.

\bibitem{Botella:2016krk}
F.~J. Botella, G.~C. Branco, M.~N. Rebelo, and J.~I. Silva-Marcos, ``{What if
  the Masses of the First Two Quark Families are not Generated by the Standard
  Higgs?},''
\href{http://arxiv.org/abs/1602.08011}{{\ttfamily arXiv:1602.08011 [hep-ph]}}.

\end{thebibliography}\endgroup

\end{document}